\documentclass[preprint,nocopyrightspace]{sigplanconf}

\usepackage{bcprules}
\usepackage{amsmath}
\usepackage{amsfonts}
\usepackage{amssymb}
\usepackage{amsthm}
\usepackage{color}
\usepackage{verbatim}
\usepackage{xspace}
\usepackage{xargs}
\usepackage{multirow}
\usepackage{graphicx}
\usepackage{booktabs}
\usepackage{stmaryrd}
\usepackage[figuresleft]{rotating}
\usepackage{galois}
\usepackage{array}
\usepackage{bm}
\usepackage{fancyvrb}
\usepackage{listings}
\usepackage{setspace}
\usepackage{url}
\usepackage{balance}
\usepackage[skip=2pt,font=small]{caption}

\newcommand{\notjs}{notJS\xspace}

\newcommand{\ljs}{\ensuremath{\lambda_{\mtt{JS}}}\xspace}

\newcommand{\etal}{et al.\xspace}

\newcommand{\stdbm}{\kw{standard}\xspace}
\newcommand{\adnbm}{\kw{addon}\xspace}
\newcommand{\emsbm}{\kw{generated}\xspace}
\newcommand{\opnbm}{\kw{opensrc}\xspace}
\newcommand{\fs}{\kw{fs}\xspace}
\newcommand{\stack}[2]{\kw{#1.#2-stack}}
\newcommand{\acyc}[1]{\kw{#1-acyclic}}
\newcommand{\objs}[2]{\kw{#1.#2-obj}}
\newcommand{\sig}[2]{\kw{#1.#2-sig}}
\newcommand{\mixed}[2]{\kw{#1.#2-mixed}}
\newcommand{\practical}{efficient\xspace}

\newcommand{\Practical}{Efficient\xspace}
\newcommand{\Practicality}{Efficiency\xspace}

\newcolumntype{R}{>{$}r<{$}}
\newcolumntype{L}{>{$}l<{$}}

\newcommand{\what}[1]{\widehat{#1}}
\newcommand{\alt}{\ \mid\ }
\newcommand{\lalt}{\ \ \ \alt}
\newcommand{\anywhere}[1]{\ensuremath{\mbox{#1}}}
\newcommand{\kw}[1]{\anywhere{\sffamily{\bfseries\small {#1}}}}
\newcommand{\red}[1]{{\color{red}{#1}}}
\newcommand{\blue}[1]{{\color{blue}{#1}}}

\newcommand{\note}[1]{\red{\textbf{[#1]}}}
\newcommand{\draft}[1]{\blue{#1}}

\newcommand{\ignore}[1]{}

\newcommand{\nvsp}{\vspace{-.1in}}
\newcommand{\cons}{\ensuremath{\!::\!}\xspace}

\newcommand{\mtt}[1]{\ensuremath{\mathit{#1}}}

\newcommand{\ttt}[1]{\ensuremath{\mbox{\small \texttt{#1}}}}

\newcommand{\Num}{\ensuremath{\mtt{Num}}\xspace}
\newcommand{\Label}{\ensuremath{\mtt{Label}}\xspace}
\newcommand{\Bool}{\ensuremath{\mtt{Bool}}\xspace}
\newcommand{\String}{\ensuremath{\mtt{String}}\xspace}
\newcommand{\Variable}{\ensuremath{\mtt{Variable}}\xspace}
\newcommand{\UnOp}{\ensuremath{\mtt{UnOp}}\xspace}
\newcommand{\BinOp}{\ensuremath{\mtt{BinOp}}\xspace}
\newcommand{\Exp}{\ensuremath{\mtt{Exp}}\xspace}
\newcommand{\Stmt}{\ensuremath{\mtt{Stmt}}\xspace}
\newcommand{\Decl}{\ensuremath{\mtt{Decl}}\xspace}

\newcommand{\Term}{\ensuremath{\mtt{Term}}\xspace}
\newcommand{\Object}{\ensuremath{\mtt{Object}}\xspace}

\newcommand{\Method}{\ensuremath{\mtt{Meth}}}
\newcommand{\Class}{\ensuremath{\mtt{Class}}}
\newcommandx{\cfg}[4][2=\env,3=\store,4=\kont,usedefault]
	{\ensuremath{\langle#1,#2,#3,#4\rangle}}
\newcommandx{\acfg}[4][2=\aenv,3=\astore,4=\akont,usedefault]
	{\ensuremath{\langle#1,#2,#3,#4\rangle}}	
\newcommand{\stmt}{\ensuremath{\mtt{s}}\xspace}
\newcommand{\expr}{\ensuremath{\mtt{e}}\xspace}
\newcommand{\variable}{\ensuremath{\mtt{x}}\xspace}

\newcommand{\kont}{\ensuremath{\kappa}}

\newcommand{\ahaltk}{\what{\kw{haltK}}\xspace}
\newcommand{\aseqk}[2]{\what{\kw{seqK}}\;\ensuremath{#1\,#2}}
\newcommand{\awhilek}[3]{\what{\kw{whileK}}\;\ensuremath{#1\,#2\,#3}}
\newcommand{\aretk}[3]{\what{\kw{retK}}\;\ensuremath{#1\,#2\,#3}}
\newcommand{\atryk}[4]{\what{\kw{tryK}}\;\ensuremath{#1\,#2\,#3\,#4}}
\newcommand{\acatchk}[2]{\what{\kw{catchK}}\;\ensuremath{#1\,#2}}
\newcommand{\afinallyk}[2]{\what{\kw{finK}}\;\ensuremath{#1\,#2}}
\newcommand{\alabelk}[2]{\what{\kw{lblK}}\;\ensuremath{#1\,#2}}
\newcommand{\afork}[4]{\what{\kw{forK}}\;\ensuremath{#1\,#2\,#3\,#4}}
\newcommand{\aaddrk}[1]{\what{\kw{addrK}}\;\ensuremath{#1}}

\newcommand{\bop}{\ensuremath{\oplus}}
\newcommand{\uop}{\ensuremath{\odot}}

\newcommand{\seq}[1]{\ensuremath{\vec{#1}_i}}
\newcommand{\bseq}[1]{\ensuremath{\overrightarrow{#1}_i}}

\newcommand{\while}[2]{\ensuremath{\kw{while }#1\; #2}}
\newcommand{\trycatch}[4]{\ensuremath{\kw{try-catch-fin }#1\, #2\, #3\, #4}}
\newcommand{\brk}[2]{\ensuremath{\kw{jump }#1 \; #2}}
\newcommand{\lbl}[2]{#1 \; #2}
\newcommand{\throw}[1]{\ensuremath{\kw{throw }#1}}

\newcommand{\cond}[3]{\ensuremath{\kw{if }#1\; #2\; #3}}
\newcommand{\str}{\ensuremath{\mtt{str}}\xspace}
\newcommand{\undefv}{\kw{undef}\xspace}
\newcommand{\nullv}{\kw{null}\xspace}

\newcommand{\true}{\kw{true}\xspace}
\newcommand{\false}{\kw{false}\xspace}
\newcommand{\block}[2]{\ensuremath{\kw{decl }#1\kw{ in }#2}}
\newcommandx{\obj}[2][2=\!]{\kw{newobj }#1\; #2}
\newcommand{\method}[1]{\ensuremath{(\self, \args) \Rightarrow #1}}
\newcommand{\self}{\ttt{self}\xspace}
\newcommand{\args}{\ttt{args}\xspace}
\newcommand{\del}[1]{\kw{del }#1\xspace}

\newcommand{\bv}{\mtt{bv}}
\newcommand{\ev}{\mtt{ev}}
\newcommand{\jv}{\mtt{jv}}

\newcommand{\tostr}{\kw{tostr }}
\newcommand{\toobj}{\kw{toobj }}
\newcommand{\tonum}{\kw{tonum }}
\newcommand{\rem}{\%}
\newcommand{\rsh}{>\!\!>}
\newcommand{\ursh}{>\!\!>\!\!>}
\newcommand{\lsh}{<\!\!<}
\newcommand{\xor}{\ensuremath{\veebar}}
\newcommand{\strcons}{+\!\!+\xspace}
\newcommand{\typeof}{\kw{typeof}}
\newcommand{\tobool}{\kw{tobool}}
\newcommand{\instance}{\kw{instanceof}}
\newcommand{\bopin}{\kw{in}}
\newcommand{\isprim}{\kw{isprim}}
\newcommand{\access}{.}

\newcommand{\Env}{\ensuremath{\mtt{Env}}\xspace}
\newcommand{\Store}{\ensuremath{\mtt{Store}}\xspace}
\newcommand{\BValue}{\ensuremath{\mtt{BValue}}\xspace}
\newcommand{\EValue}{\ensuremath{\mtt{EValue}}\xspace}
\newcommand{\JValue}{\ensuremath{\mtt{JValue}}\xspace}
\newcommand{\Value}{\ensuremath{\mtt{Value}}\xspace}
\newcommand{\Address}{\ensuremath{\mtt{Address}}\xspace}
\newcommand{\Closure}{\ensuremath{\mtt{Closure}}\xspace}
\newcommand{\CState}{\mtt{State}\xspace}

\newcommand{\state}{\ensuremath{\varsigma}\xspace}
\newcommand{\env}{\ensuremath{\rho}\xspace}
\newcommand{\store}{\ensuremath{\sigma}\xspace}

\newcommand{\clo}{\ensuremath{\mtt{clo}}\xspace}

\newcommand{\exc}[1]{\kw{exc }#1}
\newcommand{\jmp}[2]{\kw{jmp }#1\;#2}

\newcommand{\ctor}{\ensuremath{\kw{ctor}}\xspace}
\newcommand{\call}{\ensuremath{\kw{call}}\xspace}

\newcommand{\Kont}{\ensuremath{\mtt{Kont}}\xspace}
\newcommand{\mat}[1]{\ensuremath{#1}}

\newcommand{\astate}{\ensuremath{\hat{\state}}\xspace}
\newcommand{\akont}{\ensuremath{\hat{\kappa}}\xspace}

\newcommand{\astore}{\ensuremath{\hat{\store}}\xspace}
\newcommand{\abv}{\ensuremath{\widehat{\bv}}\xspace}
\newcommand{\aev}{\ensuremath{\widehat{\ev}}\xspace}
\newcommand{\ajv}{\ensuremath{\widehat{\jv}}\xspace}
\newcommand{\aenv}{\ensuremath{\hat{\env}}\xspace}
\newcommand{\av}{\ensuremath{\hat{v}}\xspace}
\newcommand{\ao}{\ensuremath{\hat{o}}\xspace}
\newcommand{\at}{\ensuremath{\hat{t}}\xspace}
\newcommand{\an}{\ensuremath{\hat{n}}\xspace}
\newcommand{\ab}{\ensuremath{\hat{b}}\xspace}

\newcommand{\astr}{\ensuremath{\what{\mtt{str}}}\xspace}
\newcommand{\abop}{\ensuremath{\hat{\bop}}\xspace}
\newcommand{\auop}{\ensuremath{\hat{\uop}}\xspace}
\newcommand{\aad}{\ensuremath{\hat{a}}\xspace}

\newcommand{\aclo}{\ensuremath{\what{\clo}}\xspace}

\newcommand{\proj}[2]{\ensuremath{\pi_{#1}(#2)}\xspace}

\newcommand{\uset}[1]{\mat{\overline{#1}}}
\newcommand{\power}{\mathcal{P}}

\newcommand{\adom}[1]{\ensuremath{#1^\sharp}\xspace}
\newcommand{\evalTo}[1]{\ensuremath{\llbracket #1 \rrbracket}}

\newcommand{\newfun}[2]{\ensuremath{\kw{newfun }#1\;#2}}
\newcommand{\newcall}[2]{\ensuremath{\kw{new }#1(#2)}}
\newcommand{\iseq}[2]{\ensuremath{\overrightarrow{#1_i = #2_i}}}

\usepackage{subfig}
\usepackage{caption}
\usepackage{tikz}
\usepackage{booktabs}
\usepackage{multirow}
\usepackage{wasysym}
\usepackage{listings}
\usepackage{enumitem}
\usepackage{algorithm}
\usepackage{algpseudocode}

\lstdefinelanguage{JavaScript}{
  morekeywords={var, function},
  morecomment=[s]{/*}{*/},%
  morecomment=[l]//,%
  morestring=[b]",%
  morestring=[b]'%
}

\newcommand{\ra}[1]{\renewcommand{\arraystretch}{#1}}

\begin{document}
\copyrightyear{2014} 
%\copyrightdata{[to be supplied]} 
%
%\preprintfooter{Permission to make digital or hard copies of all or part of this work for personal or classroom use is granted provided that copies are not made or distributed for profit or commercial advantage and that copies bear this notice and the full citation on the first page. To copy otherwise, to republish, to post on servers or to redistribute to lists, requires prior specific
%permission.}

\title{JSAI: Designing a Sound, Configurable, and \Practical Static Analyzer for JavaScript}

\authorinfo{Vineeth Kashyap$^\dagger$ \and Kyle Dewey$^\dagger$ \and Ethan A. Kuefner$^\dagger$ \and John Wagner$^\ddagger$ \\ Kevin Gibbons$^\ddagger$ \and John Sarracino$^\Upsilon$ \and Ben Wiedermann$^\Upsilon$ \and Ben Harkdekopf$^\dagger$}{University of California Santa Barbara$^\dagger$ \linebreak \{vineeth, kyledewey, eakuefner, benh\}@cs.ucsb.edu \linebreak \linebreak University of California Santa Barbara$^\ddagger$ \linebreak \{john\_wagner, kgibbons\}@umail.ucsb.edu \linebreak \linebreak Harvey Mudd College$^\Upsilon$ \linebreak \{jsarracino@g, benw@cs\}.hmc.edu}

\maketitle

\begin{abstract}

We describe JSAI, an abstract interpreter for JavaScript. JSAI uses
novel abstract domains to compute a reduced product of type inference,
pointer analysis, string analysis, integer and boolean constant
propagation, and control-flow analysis. In addition, JSAI allows for
analysis control-flow sensitivity (i.e., context-, path-, and
heap-sensitivity) to be modularly configured without requiring any
changes to the analysis implementation. JSAI is designed to be
provably sound with respect to a specific concrete semantics for
JavaScript, which has been extensively tested against existing
production-quality JavaScript implementations.

We provide a comprehensive evaluation of JSAI's performance and
precision using an extensive benchmark suite. This benchmark suite
includes real-world JavaScript applications, machine-generated
JavaScript code via Emscripten, and browser addons. We use JSAI's
configurability to evaluate a large number of analysis sensitivities
(some well-known, some novel) and observe some surprising results.  We
believe that JSAI's configurability and its formal specifications
position it as a useful research platform to experiment on novel
sensitivities, abstract domains, and client analyses for JavaScript.

\end{abstract}

\section{Introduction}
\label{sec:intro}

JavaScript is pervasive. While it began as a client-side webpage
scripting language, JavaScript is now used for a wide variety of
purposes---for example, to extend the functionality of web browsers in
the form of browser addons, to develop desktop applications (e.g., for
Windows 8~\cite{Esposito2012}) and server-side applications (e.g.,
using Node.js~\cite{nodejs}), and to develop mobile phone applications
(e.g., for Firefox OS~\cite{FirefoxOS}). A growing number of
languages, from C to Haskell, can now be compiled to
JavaScript~\cite{StarToJS}. JavaScript's growing prominence means that
secure, correct, maintainable and fast JavaScript code is becoming
ever more critical. Static analysis traditionally plays a large role
in providing these characteristics: it can be used for security
auditing, error-checking, debugging, optimization, and program
refactoring, among other uses. Thus, a sound, precise static analysis
platform for JavaScript can be of enormous advantage.

JavaScript is an inherently dynamic language: it is dynamically typed,
object properties (the JavaScript name for object members) can be
dynamically inserted and deleted, prototype-based inheritance allows
inheritance relations to be changed dynamically, implicit type conversions are abundant and can trigger user-defined code, and more. This
dynamism makes static analysis of JavaScript a significant
challenge. Compounding this difficulty is the fact that JavaScript
analysis is a relatively new endeavor---we as a community have barely
begun to explore the many possible approximations and abstractions
that balance precision and performance, in order to determine which
ones are most appropriate for JavaScript.

The current state-of-the-art static analyses for JavaScript usually
take one of two approaches: either \textbf{(1)} an
unsound\footnote{Most examples of this approach are intentionally
  unsound as a design decision, in order to handle the difficulties
  raised by JavaScript analysis. While unsound analysis can be useful
  for certain purposes, for other purposes (e.g., security auditing of
  critical code such as browser addons) sound analysis is a definite
  plus.}  dataflow analysis-based approach using baked-in data
abstractions and baked-in context- and
heap-sensitivities~\cite{Bandhakavi2011,Guha2009,Jang2009},
or \textbf{(2)} a formally-specified type system, proven sound with
respect to a specific JavaScript formal semantics but restricted to a
small subset of the full JavaScript
language~\cite{Thiemann2005,Heidegger2010,Chugh2012,Guha2011}.

In this work we introduce JSAI, the JavaScript Abstract Interpreter.
Our goal is to push the state of the art in JavaScript static analysis
along several dimensions. Specifically, our design goals for JSAI are:

\paragraph{\textit{Soundness.}} Our research question is how far we can
push sound analysis for full JavaScript while remaining practical, in contrast
with most existing full JavaScript analyses which give up on soundness due
to JavaScript's complexity. JSAI is based on the theory of abstract
interpretation~\cite{Cousot1977}, which formally relates the soundness
and precision of an abstract semantics (i.e., the static analysis)
with a given concrete semantics. Existing proposed concrete semantics
for JavaScript turn out to be inadequate for this purpose; we have
designed both concrete and abstract semantics for JavaScript
specifically for abstract interpretation. We have designed JSAI so
that its implementation closely corresponds to its formal
specification---it is, in effect, an executable semantics. JSAI
handles JavaScript as specified by the ECMA 3 standard~\cite{ECMA-262}
(sans \ttt{eval} and family; this is further discussed in
Section~\ref{sec:dynamic-code}), along with various language
extensions such as Typed Arrays~\cite{typedarray}.

\paragraph{\textit{Configurability.}} In order to explore the space of
 possible approximations and abstractions for JavaScript analysis, we
 have designed JSAI to be easily configurable in several ways. First,
 we enable context-, path- and heap-sensitivity of the analysis to be
 modularly configured without requiring changes to the rest of the analysis implementation, thus making sensitivity an independent concern. None of the existing
 static analyses for JavaScript have this capability. Doing so
 requires novel theoretical insights as detailed in the work by
 Hardekopf et. al.~\cite{Hardekopf2014}; we provide the first
 implementation of the insights contained in that paper for a
 real-world (i.e., non-toy) language. Analysis designers can specify
 known or novel sensitivities (e.g., $k$-CFA, object sensitivity, property simulation) by implementing a simple
 API that can then be plugged into the analysis. We have implemented
 over a dozen sensitivities in this manner, each of them requiring
 only 5--20 lines of code. Secondly, the string and
 number abstract domains used by the analysis are designed to be easily
 be swapped out for new, experimental abstract domains. 
 Strings and numbers are prevalent in JavaScript, and therefore designing the right abstractions for these can have a useful impact on analysis precision and performance.
 We have designed these domains each implement a specific API which is
 used by the rest of the analysis; any abstract domain implementing
 these APIs can be used in their place.

\paragraph{\textit{\Practicality.}} JSAI is designed to be competitive
with existing JavaScript analyses in terms of performance while still
meeting its goals of soundness and configurability. It incorporates
various analysis optimizations to enable it to scale to real-world
JavaScript programs of non-trivial size. JSAI is comparable in
performance to TAJS~\cite{Jensen2009, Jensen2010}, the most closely related JavaScript
analyzer, while being significantly more configurable and being based
on a formalized semantics.

The contributions of the JSAI project include complete formalisms for
a concrete and abstract semantics for JavaScript along with
implementations of concrete and abstract interpreters based on these
semantics. While concrete semantics for JavaScript have been proposed
before, ours is the first designed specifically for abstract
interpretation. Our abstract semantics is the first formal abstract
semantics for JavaScript in the literature. The abstract interpreter
implementation is the first available static analyzer for JavaScript
that provides easy configurability as a design goal. All these
contributions are available freely for download as supplementary
materials\footnote{At this URL: \url{http://cs.ucsb.edu/~vineeth/arxiv/jsai.zip}}. Thus JSAI provides a research platform to experiment with a
variety of context-, path- and heap-sensitivities and abstract
domains, and it provides a solid foundation on which to build multiple
client analyses for JavaScript. 
In fact, JSAI has been used to build a security auditing tool for browser addons~\cite{Kashyap2014}, and to experiment with type refinement as a strategy to improve analysis precision~\cite{Kashyap2013}.
The contributions of this paper
include:

\begin{itemize}
\item The design of a JavaScript intermediate language and concrete
  semantics intended specifically for abstract interpretation
  (Section~\ref{ssec:notjs}).

\item The design of an abstract semantics that enables configurable,
  sound abstract interpretation for JavaScript
  (Section~\ref{ssec:abstract}). This abstract semantics represents a
  reduced product~\cite{Cousot1979} of type inference, pointer
  analysis, string analysis, integer and boolean constant propagation,
  and control-flow analysis---all working together in
  carefully-designed harmony to enable precise tracking of data- and
  control-flow within a JavaScript program.

\item Novel abstract string and object domains for JavaScript analysis (Section~\ref{ssec:domains}).

\item Two novel context sensitivities for JavaScript
  (Section~\ref{sec:sensitivity}).

\item An evaluation of JSAI's performance and precision on the most
  comprehensive suite of benchmarks for JavaScript static analysis
  that we are aware of, including browser addons, machine-generated
  programs via Emscripten~\cite{Emscripten}, and open-source
  JavaScript programs (Section~\ref{sec:eval}). We showcase JSAI's
  configurability by evaluating a large number of context- and
  heap-sensitivities, and point out novel insights from the results.
\end{itemize}

We preface these contributions with a discussion of related work
(Section~\ref{sec:related}) \ignore{and a brief background on the
  JavaScript language (Section~\ref{sec:background}),} and conclude
with plans for future work (Section~\ref{sec:conclusion}).

\section{Related Work}
\label{sec:related}

\ignore{Other JavaScript analyses have been proposed for purposes such
  as security~\cite{Guha2009,Guarnieri2009,Chugh2009} and
  refactoring~\cite{Feldthaus2011}. These prior works have taken
  approaches ranging from fully dynamic analysis to fully static
  analysis, including hybrids of the two.} 

In this section we discuss existing static and hybrid approaches to
analyzing JavaScript, and also discuss previous efforts to formalize
JavaScript semantics. Finally, we discuss previous efforts for
configurable static analysis.

\subsection{JavaScript Analyses}

Previous work on analyzing JavaScript programs either gives up
soundness, or analyzes a restricted subset of the language, or both.
None of the previous JavaScript analyses target
configurability.\ignore{ or has studied what context- and
  heap-sensitivities work well for JavaScript.}\ignore{; the
  sensitivities they use range from \note{flow-sensitive?}
  \note{0?}-CFA~\cite{Guha2009} to flow-sensitive, 1-object
  sensitive~\cite{Jensen2009} to flow- and
  context-insensitive~\cite{Jang2009}.}

%\paragraph{\textit{JavaScript subsets.}}
Various previous work~\cite{Anderson2005, Thiemann2005, Jang2009,
  Logozzo2010, Guarnieri2009, Taly2011, Gardner2012} proposes different
subsets of the JavaScript language, and provides analyses for that
subset. These analyses range from type inference, to pointer analysis,
to numeric range and kind analysis for program optimization. None of
these handle the full complexities of JavaScript.

%\paragraph{\textit{Unsound analyses.}}
Unsound analysis can be useful under certain circumstances, and there
have been intentionally unsound analyses~\cite{Madsen2013,
  Bandhakavi2011, DoctorJS} proposed for JavaScript.  Other
works~\cite{Jang2009, Guha2009} take a best-effort approach to
soundness, without any assurance that the analysis is actually sound.

%\paragraph{\textit{Type systems.}}
Several type systems~\cite{Thiemann2005, Heidegger2010, Guha2011,
  Chugh2012} have been proposed to retrofit JavaScript (or subsets
thereof) with static types. Guha
et. al.~\cite{Guha2011} propose a novel combination of type systems
and flow analysis, and apply it to JavaScript. Chugh
et. al.~\cite{Chugh2012} propose a flow-sensitive refinement type
system designed to allow typing of common JavaScript idioms.
These type systems require programmer
annotations and cannot be used as-is on real-world JavaScript
programs, as opposed to our fully automatic approach. 

%\paragraph{\textit{Hybrid analyses.}}
Combinations of static analysis with dynamic
checks~\cite{Guarnieri2009, Chugh2009} have been proposed to handle
JavaScript---these systems statically analyze a subset of JavaScript
under certain assumptions and use runtime checks to enforce these
assumptions.  Sch\"{a}fer et al.~\cite{Schafer2013} use a dynamic
analysis to determine information that can be leveraged to scale
static analysis for JavaScript.  These ideas can usefully supplement
our static techniques.

%\paragraph{\textit{TAJS.}}
Jensen et. al.~\cite{Jensen2009} present a state-of-the-art static
analysis for JavaScript. They have since improved on this analysis in
several ways~\cite{Jensen2010, Jensen2012}; we refer to this entire
body of work as TAJS. TAJS translates JavaScript programs into a
flowgraph-based IR upon which the analysis is run. While TAJS is
intended to be sound, there is no attempt to formalize the translation
to the IR, the semantics of the IR, or the analysis itself. In fact,
while formalizing our work we found some subtle and previously unknown
soundness bugs in TAJS. TAJS also does not have the design goal of
configurability, therefore it does not allow for tunable control-flow
sensitivity or modularly replacing various abstract domains.

\subsection{JavaScript formalisms}

None of the previous work on static analysis of JavaScript has
formally specified the analysis or attempted to prove
soundness. However, there has been previous work on providing
JavaScript with a formal semantics.

%\paragraph{\textit{Direct semantics}}
Maffeis et. al~\cite{Maffeis2008} give a structural smallstep
operational semantics directly to the full JavaScript language (except
a few constructs). Lee et. al~\cite{Lee2012} propose SAFE, a semantic
framework that provides structural bigstep operational semantics to
JavaScript, based directly on the ECMAScript specification. Due to
their size and complexity, neither of these semantic formulations are
suitable for direct translation into an abstract interpreter.

%\paragraph{\textit{Core calculus approach}}
Guha et. al~\cite{Guha2010} propose a core calculus approach to
provide semantics to JavaScript---they provide a desugarer (parts of
which are formally specified) from JavaScript to a core calculus
called \ljs, which has a smallstep structural operational semantics.
Their intention was to provide a minimal core calculus that would ease
proving soundness for type systems, thus placing all the complexity in
the desugarer.  However, their core calculus is too low-level to
perform a precise and scalable static analysis (for example, some of
the semantic structure that is critical for a precise analysis is
lost, and their desugaring causes a large code bloat). We also use the
core calculus approach; however, our own intermediate language notJS is
designed to be in a sweet-spot---the complexity is shared between the
translator and the notJS semantics with the emphasis placed on static
analysis. In addition, we use an abstract machine-based semantics
rather than a structural semantics, which (as described later) is what
enables configurable analysis sensitivity.

\subsection{Configurable Analysis}

Sergey at al.~\cite{Sergey2013} describe monad-based techniques for
abstracting certain characteristics of an abstract interpreter,
allowing the analysis behavior to be configured by plugging in
different independently-specified monads. They demonstrate their
technique for lambda calculus and for Featherweight Java. However,
their work does not allow analysis sensitivity to be configured in
this way (and, in fact, their described analyses have intractable
complexity). Our work is complementary, in that we show how to make
the analysis sensitivity configurable in a manner that allows the
analysis tractability to be controlled. In addition, we demonstrate
our technique on a complete real-world language, JavaScript.

%% omitted for space
%\input{background}

\section{JSAI Design}
\label{sec:jsai}

We break our discussion of the JSAI design into three main components:
\textbf{(1)} the design of an intermediate representation for
JavaScript programs, called \notjs, along with its concrete semantics;
\textbf{(2)} the design of an abstract semantics for \notjs that
yields the reduced product of a number of essential sub-analyses and
also enables configurable analysis; and \textbf{(3)} the design of novel abstract domains for JavaScript analysis. We conclude with a
discussion of various options for handling dynamic code injection.

The intent of this section is to discuss the design decisions that
went into JSAI, rather than giving a comprehensive description of the
various formalisms (e.g., the translation from JavaScript to \notjs,
the concrete semantics of \notjs, and the abstract semantics of
\notjs). All of these formalisms, along with their implementations,
appear in the supplementary materials.

\subsection{Designing the \notjs Intermediate Language}
\label{ssec:notjs}

Our soundness goal motivates the use of formal specifications for both
concrete JavaScript semantics and our abstract analysis semantics. Our
approach is to define an intermediate language called \notjs, along
with a formally-specified translation from JavaScript to \notjs. We
then give \notjs a formal concrete semantics upon which we base our
abstract interpreter.\footnote{Guha et al~\cite{Guha2010} use a similar
  approach, but our IR design and formal semantics are different. See
  Section~\ref{sec:related} for a discussion of the differences
  between our two approaches.}

\ignore{ The choice we faced was whether to \textbf{(1)} provide
  formalisms directly for the JavaScript language itself; or
  \textbf{(2)} translate JavaScript into an intermediate
  representation and provide formalisms for that. Several previous
  works have provided semantics directly for the JavaScript
  language~\cite{Lee2012, Maffeis2008}, adhering as closely as
  possible to the ECMA standard. The lesson we learn from those
  efforts is that the resulting formalisms are large, complex, and
  difficult to reason about; abstracting those semantics would also be
  difficult and inefficient. Therefore, we chose the other option by
  designing an intermediate language called \notjs, along with a
  formally-specified translation from JavaScript to
  \notjs \footnote{Guha et al~\cite{Guha2010} use a similar approach,
    but our IR design is different. See Section~\ref{sec:related} for
    a discussion of the differences between the two approaches.}. The
  semantics of \notjs is much smaller and simpler than the other
  approach, and hence easier to formalize and implement. }

Figure~\ref{fig:syntax} shows the abstract syntax of \notjs, which was
carefully designed with the ultimate goal of making abstract
interpretation simple, precise, and efficient. JavaScript's builtin
objects (e.g,. \lstinline|Math|) and methods (e.g., \lstinline|isNaN|)  
are properties of the global object constructed prior to a program's
execution, thus they are not a part of the language syntax. 

\begin{figure}[!ht]
\small
\begin{gather*}
  n \in \Num \quad b \in \Bool \quad \str \in \String \quad
  \variable \in \Variable \quad \ell \in \Label
\end{gather*}
\nvsp\nvsp
\begin{align*}
  \stmt \in \Stmt &::= \seq{\stmt} \alt \cond{{\expr}}{\stmt_1}{\stmt_2} \alt \while{{\expr}}{\stmt} \alt \variable := e \alt \expr_1.\expr_2 := \expr_3 \\
  &\lalt \variable := \expr_1(\expr_2,\expr_3) \alt \variable := \toobj e \alt
  \variable := \del{\expr_1.\expr_2} \\
  &\lalt \variable := \newfun{m}{n} \alt \variable := \newcall{\expr_1}{\expr_2} \alt \throw{{\expr}} \\
  &\lalt \trycatch{\stmt_1}{x}{\stmt_2}{\stmt_3} \alt \lbl{\ell}{\stmt} \alt \brk{\ell}{{\expr}} \alt \kw{for }x\,e\,s
  \\
  \expr \in \Exp &::= n \alt b \alt \str \alt \undefv \alt \nullv 
  \alt x \alt m \alt \expr_1 \bop \expr_2 \alt \uop e \\
  d \in \Decl &::= \block{\iseq{x}{{\expr}}}{\stmt}
  \\
  m \in \Method &::= \method{d} \alt \method{\stmt}
  \\
  \bop \in \BinOp &::= + \alt - \alt \times \alt \div \alt \rem \alt
  \lsh \alt \rsh \alt \ursh \alt < \alt \leq \alt \& \\
  &\lalt '|' \alt \xor \alt \kw{and} \alt \kw{or} \alt \strcons \alt
  \prec \alt \preceq \alt \approx \alt \equiv \alt \access \\
  &\lalt \instance \alt \bopin 
  \\
  \uop \in \UnOp &::= - \alt \sim\; \alt \neg \alt \typeof \alt
  \isprim \alt \tobool \\
  &\lalt \tostr \alt \tonum
\end{align*}
\caption{\label{fig:syntax} The abstract syntax of \notjs provides
  canonical constructs that simplify JavaScript's behavior. The vector
  notation represents (by abuse of notation) an ordered sequence of
  unspecified length $n$, where $i$ ranges from 0 to $n-1$.}

\end{figure}

An important design decision we made is to separate the language into
pure expressions ($e \in \Exp$) that are guaranteed to terminate
without throwing an exception, and impure statements ($s \in \Stmt$)
that do not have these guarantees. This decision directly impacts the
formal semantics and implementation of \notjs, a further discussion of
which appears later in this section. This is the first IR for
JavaScript we are aware of that makes this design choice---it is a
more radical choice than might first be apparent, because JavaScript's
implicit conversions make it difficult to enforce this separation. The
IR was carefully designed to make this possible. Some other design
decisions of note include making JavaScript's implicit conversions
(which are complex and difficult to reason about, involving multiple
steps and alternatives depending on the current state of the program)
explicit in \notjs; leaving certain JavaScript constructs unlowered to
allow for a more precise abstract semantics (e.g., the
\ttt{for}..\ttt{in} loop, which we leave mostly intact as $\kw{for
}x\,e\,s$); and simplifying method calls to make the implicit
\ttt{this} parameter and \ttt{arguments} object explicit---\ttt{this}
is often, but not always, the address of a method's receiver object,
and its value can be non-intuitive, while \ttt{arguments} provides a
form of reflection providing access to a method's arguments.

Given the \notjs abstract syntax, we need to design a formal concrete
semantics that (together with the translation to \notjs) captures
JavaScript behavior. We have two main criteria: \textbf{(1)} the
semantics should be specified in a manner that can be directly
converted into an implementation, allowing us to test its behavior
against actual JavaScript implementations; \textbf{(2)} looking ahead
to the abstract version of the semantics (which defines our analysis),
the semantics should be specified in a manner that allows for
configurable sensitivity. These requirements lead us to specify the
\notjs semantics as an abstract machine-based smallstep operational
semantics. One can think of this semantics as an infinite state
transition system, wherein we formally define a notion of
\textit{state} and a set of \textit{transition rules} that connect
states. The semantics is implemented by turning the state definition
into a data structure (e.g., a Scala class) and the transition rules into
functions that transform a given state into the next state. The
concrete interpreter starts with an initial state (containing the
start of the program and all of the builtin JavaScript methods and
objects), and continually computes the next state until the program
finishes.

\paragraph{\textit{Further Design Discussion.}} Previous efforts to
give JavaScript a formal concrete semantics all use either bigstep or
smallstep structural operational semantics. However, a smallstep
abstract machine semantics is more suited for abstract interpretation
(particularly to enable configurability in the form of tunable
control-flow sensitivity and straightforward implementation). Our
semantics is actually not completely smallstep: expressions are
evaluated in a bigstep style, which means they are evaluated via a
recursive traversal of their abstract syntax tree (AST), similar to
most AST-based interpreters. This is made possible by our separation
of expressions (pure, terminating) from statements (impure,
potentially non-terminating). While this separation might be standard
for simpler languages, it took careful design of the \notjs IR to
enable this separation for JavaScript.

Initially we designed \notjs so that there was no separation between
statements and expressions, and side-effects and exceptions could
happen anywhere. We designed the corresponding semantics to be in
completely smallstep style. As opposed to our current design (which
keeps expressions separate and guarantees they are pure), the initial
design had three times as many semantic continuations, and more
complicated reasoning for the semantic rules.

We omit further details of the concrete semantics both for space and
because they are almost redundant with the abstract semantics
described in the next section. The main difference between the two is
that the abstract state employs sets in places where the concrete
state employs singletons, and the abstract transition rules are
nondeterministic whereas the concrete rules are deterministic. Both of
these differences are because the abstract semantics over-approximates
the concrete semantics.

\paragraph{\textit{Testing the Semantics.}}
We tested the translation, semantics, and implementation thereof by
comparing its behavior with that of an actual JavaScript engine,
SpiderMonkey~\cite{spidermonkey}. We constructed a test suite of over a million JavaScript programs, most of which were randomly generated. 
However, 243 of the programs in the test suite were either hand-crafted to exercise various parts of
the semantics, or taken from existing JavaScript programs used to
test commercial JavaScript implementations. We then ran all of the tests on SpiderMonkey and on
our concrete interpreter, and we verified that they produce identical
output. While we can never completely guarantee that the \notjs
semantics matches the ECMA specification, we can do as well as any
JavaScript implementation, which goes through the same sort of testing
process.

\subsection{Designing the Abstract Semantics}
\label{ssec:abstract}

The JavaScript static analysis is defined as an abstract semantics for
\notjs that over-approximates the \notjs concrete semantics. The
analysis is implemented by computing the set of all abstract states
reachable from a given initial state by following the abstract
transition rules. The analysis contains some special machinery that
provides configurable sensitivity. We illustrate our approach via a
worklist algorithm that ties these concepts together:

\begin{algorithm}
\caption{The JSAI worklist algorithm}\label{alg:worklist}
\algrenewcommand{\algorithmiccomment}[1]
{\\ \hskip1.5em{\textit{// #1}}}
\begin{algorithmic}[1]
\item put the initial abstract state $\astate_0$ on the worklist
\item initialize map \ttt{partition} : \mtt{Trace} $\to \adom{\CState}$ to empty
\Repeat
  \State remove an abstract state $\astate$ from the worklist 
  \ForAll{abstract states $\astate'$ in \ttt{next\_states}(\astate)}
    \If{\ttt{partition} does not contain \ttt{trace}($\astate'$)}
      \State \ttt{partition}(\ttt{trace}($\astate'$)) = $\astate'$
      \State put $\astate'$ on worklist
    \Else
      \State $\astate_{\mtt{old}}$ = \ttt{partition}(\ttt{trace}($\astate'$))
      \State $\astate_{\mtt{new}}$ = $\astate_{\mtt{old}} \sqcup \astate'$
      \If{$\astate_{\mtt{new}} \neq \astate_{\mtt{old}}$}
        \State \ttt{partition}(\ttt{trace}($\astate'$)) = $\astate_{\mtt{new}}$
        \State put $\astate_{\mtt{new}}$ on worklist
      \EndIf
    \EndIf
  \EndFor
\Until{worklist is empty}
\end{algorithmic}
\end{algorithm}

The static analysis performed by this worklist algorithm is determined
by the definitions of the abstract semantic states $\astate \in
\adom{\CState}$, the abstract transition rules
$\ttt{next\_states}(\astate) \in \adom{\CState} \to
\power(\adom{\CState})$, and the knob that configures the analysis
sensitivity $\ttt{trace}(\astate)$. We discuss each of these aspects
in turn.

\paragraph{\textit{Abstract Semantic Domains.}}
Figure~\ref{fig:adomains} shows our definition of an abstract state
for \notjs. An abstract state $\astate$ consists of a \textit{term}
that is either a \notjs statement or an abstract value that is the
result of evaluating a statement; an \textit{environment} that maps
variables to (sets of) addresses; a \textit{store} mapping addresses
to either abstract values, abstract objects, or sets of continuations
(to enforce computability for abstract semantics that use semantic
continuations, as per Van Horn and Might~\cite{VanHorn2010}); and
finally a \textit{continuation stack} that represents the remaining
computations to perform---one can think of this component as analogous
to a runtime stack that remembers computations that should completed
once the current computation is finished.

\begin{figure}
\small
\centering
\begin{gather*}
  \an \in \adom{\Num} \quad \astr \in \adom{\String} \quad \aad \in
  \adom{\Address} \quad \auop \in \adom{\UnOp} \quad \abop \in
  \adom{\BinOp}
\end{gather*} 
{\centering \begin{align*} 
    \astate \in \adom{\CState} &=
    \adom{\Term} \times \adom{\Env} \times \adom{\Store} \times
    \adom{\Kont}
    \\
    \at \in \adom{\Term} &= \Decl + \Stmt + \adom{\Value}
    \\
    \aenv \in \adom{\Env} &= \Variable \rightarrow
    \power(\adom{\Address})
    \\
    \astore \in \adom{\Store} &= \adom{\Address} \rightarrow
    (\adom{\BValue} + \adom{\Object} + \power(\adom{\Kont}))
    \\
    \abv \in \adom{\BValue} &= \adom{\Num} \times \power(\Bool) \times
    \adom{\String} \times \power(\adom{\Address}) \times \\
    &\;\quad \power(\{\nullv\}) \times \power(\{\undefv\})
    \\
    \ao \in \adom{\Object} &= (\adom{\String} \to \adom{\BValue}) \times \power(\String)
    \times \\
    &\;\quad(\String \to (\adom{\BValue} + \Class + \power(\adom{\Closure})))  \\
    c \in \Class &= \{ \kw{function}, \kw{array}, \kw{string},
    \kw{boolean}, \kw{number}, \kw{date}, \\
    &\qquad \kw{error}, \kw{regexp},
    \kw{arguments}, \kw{object}, \ldots \}
    \\
    \aclo \in \adom{\Closure} &= \adom{\Env} \times \Method
    \\
    \aev \in \adom{\EValue} &::= \exc{\bv}
    \\
    \ajv \in \adom{\JValue} &::= \jmp \ell \abv
    \\
    \av \in \adom{\Value} &= \adom{\BValue} + \adom{\EValue} +
    \adom{\JValue}
    \\
    \akont \in \adom{\Kont} &::= \ahaltk \alt \aseqk{\seq{\stmt}}{\akont} \alt \awhilek{\expr}{\stmt}{\akont} \alt \alabelk{\ell}{\akont} \\
    & \lalt \afork{\bseq{\astr}}{\variable}{\stmt}{\akont} \alt \aretk{\variable}{\aenv}{\akont}\ {\ctor} \alt \aretk{\variable}{\aenv}{\akont}\ {\call}\ \\
    & \lalt \atryk{\variable}{\stmt}{\stmt}{\akont} \alt \acatchk{\stmt}{\akont} \alt \afinallyk{\uset{\av}}{\akont} \alt \aaddrk{\aad}
  \end{align*}
}
\caption{Abstract semantic domains for \notjs.}
\label{fig:adomains}
\end{figure}

\ignore{
  \begin{itemize}
  \item[] \textbf{a Term} that is either a \notjs statement,
    signifying the next statement to process, or an abstract value
    representing the result of processing a statement.

  \item[] \textbf{an Environment} that maps variable names to (sets
    of) abstract addresses, enforcing lexical scope.

  \item[] \textbf{a Store} that maps abstract addresses to either
    abstract values, abstract objects, or sets of continuations. The
    latter is used to enforce computability for abstract semantics
    that use semantic continuations, as per Van Horn and
    Might~\cite{VanHorn2010}.

  \item[] \textbf{a Continuation Stack} that represents the remaining
    computations to perform; one can think of it as an analogue to a
    runtime stack that remembers computations to be completed once the
    current computation is finished. Mathematically the continuation
    stack is specified similarly to defining a list using \kw{cons},
    where each element contains a reference to the next element.
  \end{itemize}
}

Abstract values are either exception/jump values (\adom{\EValue},
\adom{\JValue}), used to handle non-local control-flow, or base values
(\adom{\BValue}), used to represent JavaScript values. Base values are
a tuple of abstract numbers, booleans, strings, addresses, null, and
undefined; each of these components is a lattice. Base values are defined as tuples because the analysis
over-approximates the concrete semantics, and thus cannot constrain
values to be only a single type at a time. These value tuples yield a
type inference analysis: any component of this tuple that is a lattice $\bot$
represents a type that this value cannot contain. Base values do not
include function closures, because functions in JavaScript are
actually objects. Instead, we define a class of abstract objects that
correspond to functions and that contain a set of closures that are
used when that object is called as a function. We describe our novel
abstract object domain in more detail in Section~\ref{ssec:domains}.

Each component of the tuple also represents an individual analysis:
the abstract number domain determines a number analysis, the abstract
string domain determines a string analysis, the abstract addresses
domain determines a pointer analysis, etc. Composing the individual
analyses represented by the components of the value tuple is not a
trivial task; a simple cartesian product of these domains (which
corresponds to running each analysis independently, without using
information from the other analyses) would be imprecise to the point
of being useless. Instead, we specify a reduced
product~\cite{Cousot1979} of the individual analyses, which means that
we define the semantics so that each individual domain can take
advantage of the other domains' information to improve their
results. The abstract number and string domains are intentionally
unspecified in the semantics; they are configurable. We discuss our
specific implementations of the abstract string domain in
Section~\ref{ssec:domains}.

Together, all of these abstract domains define a set of simultaneous
analyses: control-flow analysis (for each call-site, which methods may
be called), pointer analysis (for each object reference, which objects
may be accessed), type inference (for each value, can it be a number,
a boolean, a string, \kw{null}, \kw{undef}, or a particular class of
object), and extended versions of boolean, number, and string constant
propagation (for each boolean, number and string value, is it a known
constant value). These analyses combine to give detailed control- and
data-flow information forming a fundamental analysis that can be used
by many possible clients (e.g., error detection, program slicing,
secure information flow).

\paragraph{\textit{Abstract Transition Rules.}}
Figure~\ref{fig:arules} describes a small subset of the abstract
transition rules, to give their flavor. To compute
$\ttt{next\_states}(\astate)$, the components of $\astate$ are matched
against the premises of the rules to find which rule(s) are relevant;
that rule then describes the next state (if multiple rules apply, then
there will be multiple next states). The rules \textsc{1}, \textsc{2},
and \textsc{3} deal with sequences of statements. Rule \textsc{1} says
that if the state's term is a sequence, then pick the first statement
in the sequence to be the next state's term; then take the rest of the
sequence and put it in a \kw{seqK} continuation for the next state,
pushing it on top of the continuation stack. Rule \textsc{2} says that
if the state's term is a base value (and hence we have completed the
evaluation of a statement), take the next statement from the \kw{seqK}
continuation and make it the term for the next state. Rule \textsc{3}
says that if there are no more statements in the sequence, pop the
\kw{seqK} continuation off of the continuation stack. The rules
\textsc{4} and \textsc{5} deal with conditionals. Rule \textsc{4} says
that if the guard expression evaluates to an abstract value that
over-approximates \kw{true}, make the \kw{true} branch statement the
term for the next state; rule \textsc{5} is similar except it takes
the \kw{false} branch. Note that these rules are nondeterministic, in
that the same state can match both rules.

\begin{figure}
\small
\centering
\ra{1.3}

\begin{tabular}{lll}\toprule
& Current State \astate & Next State $\astate'$ \\
\midrule
\textsc{1} & \acfg{s\cons\seq{s}} & \acfg{s}[][][\aseqk{\seq{s}}{\akont}]\\
\textsc{2} &\acfg{\abv}[][][\aseqk{s\cons \vec s_i}{\akont}] & \acfg{s}[][][\aseqk{\vec s_i}{\akont}]\\
\textsc{3} &\acfg{\abv}[][][\aseqk{\epsilon}{\akont}] & \acfg{\abv}[][][\akont]\\
\textsc{4} &\acfg{\cond{e}{s_1}{s_2}} & \acfg{s_1}
                  \quad {\footnotesize 
                          \emph{if} $\true \in \proj{\ab}{\evalTo{e}}$}\\
\textsc{5} &\acfg{\cond{e}{s_1}{s_2}} & \acfg{s_2}
                  \quad {\footnotesize 
                          \emph{if} $\false \in \proj{\ab}{\evalTo{e}}$}\\
\bottomrule
\end{tabular}

\caption{A small subset of the abstract semantics rules for JSAI. Each
  smallstep rule describes a transition relation from one abstract
  state \state to the next state $\astate'$. The phrase
  \proj{\ab}{\evalTo{e}} means to evaluate expression $e$ to an
  abstract base value, then project out its boolean component.}

\label{fig:arules}
\end{figure}

%The \textsc{IF*} rules use the evaluator $\evalTo{.}$ to reduce a
%pure expression to an abstract base value. To compute object property
%lookups, the evaluator uses a helper
%function. Figure~\ref{fig:clookup} gives the concrete version of
%property lookup, and Figure~\ref{fig:alook} gives the abstract
%version. The abstract property lookup must deal with nondeterminism
%when either the property name is unknown, or it is unknown whether
%the object in question contains that property. This is one example
%of a number of such helper functions used by the abstract transition
%rules, and fully specified in the semantic formalisms.

\ignore{
  \input{figures/clookup.tex}
  \input{figures/alookup.tex}
}

\paragraph{\textit{Configurable Sensitivity.}}
\label{sssec:tune}
To enable configurable sensitivity, we build on the insights of
Hardekopf et al~\cite{Hardekopf2014}. We extend the abstract state to
include an additional component from a \textit{Trace} abstract
domain. The worklist algorithm uses the \ttt{trace} function to map
each abstract state to its trace, and joins together all reachable
abstract states that map to the same trace (see lines 10--11 of
Algorithm~\ref{alg:worklist}). The definition of \textit{Trace} is
left to the analysis designer; different definitions yield different
sensitivities. For example, suppose \mtt{Trace} is defined as the set
of program points, and an individual state's trace is the current
program point. Then our worklist algorithm computes a flow-sensitive,
context-insensitive analysis: all states at the same program point are
joined together, yielding one state per program point. Suppose we
redefine \mtt{Trace} to be sequences of program points, and an
individual state's trace to be the last $k$ call-sites. Then our
worklist algorithm computes a flow-sensitive, $k$-CFA
context-sensitive analysis. Arbitrary sensitivities can be defined in
this manner solely by redefining \mtt{Trace}, without affecting the
worklist algorithm or the abstract transition rules. We explore a
number of possibilities in Section~\ref{sec:eval}.

While most static analyses are built on top of the control-flow graph
(CFG), JSAI instead employs semantic continuations inside the abstract
states. One reason is so we can employ the techniques of
\cite{Hardekopf2014} to enable configurable sensitivity. These techniques are also
an important reason that we used abstract machine-based semantics
instead of structural semantics as in all previous formalizations of
JavaScript---a structural semantics would not allow us to configure
the sensitivity in this way. Another reason for using semantic
continuations is that JavaScript contains a great deal of indirect and non-local
control-flow (due to higher-order functions, implicit exceptions,
etc), thus much of the control-flow is non-obvious and basing the
analysis on a CFG would require a great deal of ad-hoc patches during
the analysis. One class of soundness bugs relating to \ttt{try-catch-finally} that we found in TAJS, which does use a CFG, was due to exactly this issue.

\subsection{Novel Abstract Domains}
\label{ssec:domains}

JSAI allows configurable abstract number and string domains, but we
also provide default domains based on our experience with JavaScript
analysis. We motivate and describe our default abstract string domain
here. We also describe our novel abstract object domain, which is an
integral part of the JSAI abstract semantics.

\paragraph{\textit{Abstract Strings.}}
Our initial abstract string domain \adom{\String} was a simple string
constant domain---the elements were either constant strings or $\top$,
representing an unknown string. We extended that domain to separate
unknown strings into two categories: strings that are definitely
numbers, and strings that are definitely not numbers (borrowing from
TAJS~\cite{Jensen2009}). This separation helps when analyzing arrays:
arrays are just objects where array indices are represented with
numeric string properties such as \ttt{"0"}, \ttt{"1"}, etc, but they
also have non-numeric properties like \ttt{"length"}. However, this
initial string domain was inadequate.

In particular, we discovered a need to express that a string is
\textit{not} contained within a given hard-coded set of strings.
Consider the property lookup \ttt{x := obj[y]}, where \ttt{y} is a
variable that resolves to an unknown string. Because the string is
unknown, the analysis is forced to assign to \ttt{x} not only the
lattice join of all values contained in \ttt{obj}, but also the
lattice join of all the values contained in all prototypes of
\ttt{obj}, due to the rules of prototype-based inheritance. Almost all
object prototype chains terminate in one of the builtin objects
contained in the global object (\ttt{Object.prototype},
\ttt{Array.prototype}, etc); these builtin objects contain the builtin
values and methods. Thus, all of these builtin values and methods are
returned for any object property access based on an unknown string,
severely polluting the results. One possible way to mitigate this
problem is to use an expensive domain that can express arbitrary
complements (i.e., express that a string is \textit{not} contained in
some arbitrary set of strings). Instead, we extend the string domain
to separate out \textit{special} strings (\ttt{valueOf},
\ttt{toString} etc.) from the rest; these special strings are drawn
from property names of builtin values and methods. We can thus express
that a string has an unknown value that is \textit{not} one of the
special values. This is a practical solution that improves precision
at minimal cost.

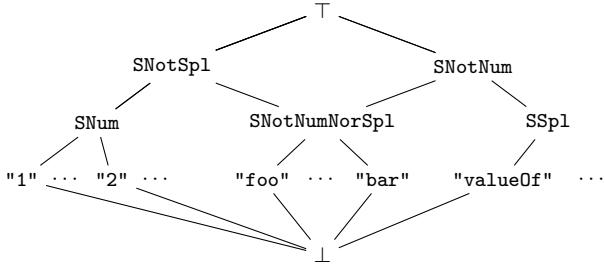
\begin{figure}
  \centering
%   \begin{subfigure}{.4\textwidth}
%   \centering
% \begin{tikzpicture}
% 	\node (top) at (0,2) {$\top$};
% 	\node (minfty) at (-2.5, 0) {$-\infty$};
% 	\node (pinfty) at (-1.5, 0) {$\infty$};
% 	\node (nan) at (-0.5,0) {\texttt{NaN}};

% 	\node (dots0) at (0.6,0) {$\cdots$};
% 	\node (n1) at (1.3,0) {$-1.5$};
% 	\node (dots1) at (2.1,0) {$\cdots$};
% 	\node (n3) at (2.9,0) {$1.5$};
% 	\node (dots3) at (3.5,0) {$\cdots$};
	
% 	\node (real) at (2.1,1) {\texttt{Real}};

% 	\node (bot) at (0,-1) {$\bot$};
% 	\draw (bot) -- (minfty) -- (top);
% 	\draw (bot) -- (pinfty) -- (top);
% 	\draw (bot) -- (nan) -- (top);
% 	\draw (bot) -- (n1) -- (real) -- (top);
% 	\draw (bot) -- (n3) -- (real) -- (top);
% \end{tikzpicture}
%   \end{subfigure}
  % \begin{subfigure}{.4\textwidth}
  % \centering
\begin{tikzpicture}
  \footnotesize
  \tikzset{my node/.style={node distance=.8cm,shape=rectangle}}

	\node (top) at (0,2) {$\top$};

  \node (snotspl) at (-2, 1.25) {\texttt{SNotSpl}};
  \node (snotnum) at (2, 1.25) {\texttt{SNotNum}};

  \node (snum) at (-3, 0.5) {\texttt{SNum}};
  \node (snotnumnorspl) at (0, 0.5) {\texttt{SNotNumNorSpl}};
  \node (sspl) at (3, 0.5) {\texttt{SSpl}};
  
  \node (one) at (-4, -0.25) {\texttt{"1"}};
  \node (snumdots) at (-3.4, -0.25) {$\cdots$};
  \node (two) at (-2.8, -0.25) {\texttt{"2"}};
  \node (snumdotstwo) at (-2.2, -0.25) {$\cdots$};

  \node (foo) at (-0.8, -0.25) {\texttt{"foo"}};
  \node (bar) at (0.8, -0.25) {\texttt{"bar"}};
  \node (snnnsdots) at (0, -0.25) {$\cdots$};

  \node (valueof) at (2.4,-0.25) {\texttt{"valueOf"}};
  \node (sspldots) at (3.6,-0.25) {$\cdots$};

	\node (bot) at (0,-1.25) {$\bot$};
	
	\draw (bot) -- (one) -- (snum) -- (snotspl) -- (top);
	\draw (bot) -- (two) -- (snum) -- (snotspl) -- (top);
  \draw (bot) -- (foo) -- (snotnumnorspl);
  \draw (bot) -- (bar) -- (snotnumnorspl);
  \draw (bot) -- (valueof) -- (sspl) -- (snotnum) -- (top);
  \draw (snotnumnorspl) -- (snotspl) -- (top);
  \draw (snotnumnorspl) -- (snotnum) -- (top);
\end{tikzpicture}
  % \end{subfigure}
  \caption{Our default string abstract domain, further explained in Section~\ref{ssec:domains}.}
  \label{fig:jsai0}
\end{figure}

The new abstract string domain depicted in Figure~\ref{fig:jsai0} (that separates unknown strings into
numeric, non-numeric and special strings) was simple to implement due
to JSAI's configurable architecture; it did not require changes to any
other parts of the implementation despite the pervasive use of strings
in all aspects of JavaScript semantics.

\paragraph{\textit{Abstract Objects.}}
We highlight the abstract domain \adom{\Object} of
Figure~\ref{fig:adomains} as a novel contribution. Previous JavaScript
analyses model abstract objects as a tuple containing \textbf{(1)} a
map from property names to values; and \textbf{(2)} a list of
definitely present properties (necessary because property names are
just strings, and objects can be modified using unknown strings as
property names). However, according to the ECMA standard objects can
be of different \textit{classes}, such as functions, arrays, dates,
regexps, etc. While these are all objects and share many similarities,
there are semantic differences between objects of different
classes. For example, the \ttt{length} property of array objects has
semantic significance---assigning a value to \ttt{length} can
implicitly add or delete properties to the array object, and certain
values cannot be assigned to \ttt{length} without raising a runtime
exception. Non-array objects can also have a \ttt{length} field, but
assigning to that field will have no other effect. The object's class
dictates the semantics of property enumerate, update and delete
operations on an object. Thus, the analysis must track what classes an
abstract object may belong to in order to accurately model these
semantic differences. If abstract objects can belong to arbitrary sets
of classes, this tracking and modeling becomes extremely complex,
error-prone, and inefficient.

Our innovation is to add a map as the third component of abstract
objects that contains class-specific values. This component also
records which class an abstract object belongs to. Finally, the
semantics is designed so that any given abstract object must belong to
exactly one class. This is enforced by assigning abstract addresses to
objects based not just on their static allocation site and context,
but also on the constructor used to create the object (which
determines its class). The resulting abstract semantics is much
simpler, more efficient, and precise.

\subsection{Handling \ttt{eval} and Similar Constructs}
\label{sec:dynamic-code}

Dynamically injected code is the bane of static analysis. JavaScript
contains \ttt{eval}, which executes an arbitrary string as
code.\footnote{There are related constructs with similar
  functionality; we refer to all of them as \ttt{eval} for
  convenience.} The \notjs IR does not contain an explicit
\ttt{eval} instruction because \ttt{eval} is a builtin
method of the global object, rather than being a JavaScript
instruction. 

There are several possible strategies to handle \ttt{eval} for static
analysis. For example, we could \textit{disallow} the use of
\ttt{eval} altogether. In some application domains this is a
legitimate strategy---e.g., browser addons must pass through a vetting
process to be added to official repositories, and this process
strongly discourages \ttt{eval}; also, machine-generated JavaScript a
la Emscripten~\cite{Emscripten} rarely contains \ttt{eval}. There are also
methods to automatically~\cite{Jensen2012} or
semi-automatically~\cite{Meawad2012} \textit{eliminate} \ttt{eval} in
most real-world scenarios. Alternatively, the analysis can make
assumptions about the runtime behavior of the \ttt{eval} statement,
and the program or runtime can be modified to \textit{check} or
\textit{enforce} these assumptions, e.g., by running \ttt{eval} inside
a sandbox. Such runtime checks are used by the staged analysis
proposed by Chugh \etal~\cite{Chugh2009}. Finally, the static analysis
can initially ignore dynamic code injection, and the runtime can be
modified to have the analysis \textit{patch} itself to soundly handle
the newly-available information; this is similar to the strategy
proposed for handling Java analysis in the presence of dynamic class
loading~\cite{Ishizaki2000}.

In this work we do not innovate on methods for handling \ttt{eval}; we
simply use the \textit{disallow} strategy and have the analysis output
a warning if \ttt{eval} could potentially have been called. For an
important and growing class of JavaScript programs, e.g., browser
addons and machine-generated programs, the \textit{disallow} strategy
is a sensible choice---none of the 28 real-world benchmarks in our
benchmark suite use \ttt{eval}. More comprehensive methods for
handling \ttt{eval} are complementary to JSAI and can be incorporated
without significant modifications to the current design.

\section{Showcasing JSAI's Configurability}
\label{sec:sensitivity}

The primary motivation for making JSAI configurable is to allow
analysis designers to explore different possibilities for
approximation and abstraction. One important dimension to explore is
\textit{context-sensitivity}: how the (potentially infinite) possible
method call instances are partitioned and merged into a finite number
of abstract instances. The context-sensitivity strategy used by an
analysis can greatly influence the analysis precision and
performance. The current state of the art for JavaScript static
analysis has explored only a few possible context-sensitivity
strategies, all of which are baked into the analysis and difficult to
change.

We take advantage of JSAI's configurability to define and evaluate a
much larger selection of context-sensitivities than has ever been
evaluated before in a single paper. Because of JSAI's design,
specifying each sensitivity takes only 5--20 lines of code, making
this process substantially more feasible than existing analysis
implementations (where each sensitivity would have to be hard-coded
into the analysis from scratch). The analysis designer specifies a
sensitivity by instantiating a particular instance of \mtt{Trace}; all
abstract states with the same trace will be merged together. For
context-sensitivity, we define \mtt{Trace} to include some notion of
the calling context, so that states in the same context are merged
while states in different contexts are kept separate.

We implement six main context-sensitivity strategies, each
parameterized in various ways, yielding a total of 56 different forms
of context-sensitivity. All of our sensitivities are flow-sensitive
(JavaScript's dynamic nature means that flow-insensitive analyses tend
to have terrible precision). We empirically evaluate all of these
strategies in Section~\ref{sec:eval}; here we define the six main
strategies. Four of the six strategies are known in the literature,
while two are novel to this paper. The novel strategies are based on
two hypotheses about context definitions that might provide a good
balance between precision and performance. Our empirical evaluation
demonstrates that these hypotheses are false, i.e., they do not
provide any substantial benefit. We include them here not as examples
of good sensitivities to use, but rather to demonstrate that JSAI
makes it easy to formulate and test hypotheses about analysis
strategies---each novel strategy took only 15--20 minutes to
implement. The strategies we defined are as follows, where the first
four are known and the last two are novel:

\paragraph{\textit{Context-insensitive.}} All calls to a given method
are merged. We define the context component of \mtt{Trace} to be a
unit value, so that all contexts are the same.

\paragraph{\textit{Stack-CFA.}} Contexts are distinguished by the list of
call-sites on the call-stack. This strategy is $k$-limited to ensure
there are only a finite number of possible contexts. We define the
\mtt{Trace} component to contain the top $k$ call-sites.

\paragraph{\textit{Acyclic-CFA.}} Contexts are distinguished the same as
Stack-CFA, but instead of $k$-limiting we collapse recursive call
cycles. We define \mtt{Trace} to contain all call-sites on the
call-stack, except that cycles are collapsed.

\paragraph{\textit{Object-sensitive.}} Contexts are
distinguished by a list of addresses corresponding to the chain of
receiver objects (corresponding to full-object-sensitivity in Smaragdakis \etal~\cite{Smaragdakis2011}). We define \mtt{Trace} to contain this information
($k$-limited to ensure finite contexts).

\paragraph{\textit{Signature-CFA.}} Type information is important for
dynamically-typed languages, so intuitively it seems that type
information would make good contexts. We hypothesize that defining
\mtt{Trace} to record the types of a call's arguments would be a good
context-sensitivity, so that all calls with the same types of
arguments would be merged.

\paragraph{\textit{Mixed-CFA.}} Object-sensitivity uses the address of
the receiver object. However, in JavaScript the receiver object is
often the global object created at the beginning of the program
execution. Intuitively, it seems this would mean that object
sensitivity might merge many calls that should be kept separate. We
hypothesized that it might be beneficial to define \mtt{Trace}
as a modified object-sensitive strategy---when object-sensitivity
would use the address of the global object, this strategy uses the
current call-site instead.

\section{Evaluation}
\label{sec:eval}

In this section we evaluate JSAI's precision and performance for a
range of context-sensitivities as described in
Section~\ref{sec:sensitivity}, for a total of 56 distinct
sensitivities. We run each sensitivity on 28 benchmarks collected from
four different application domains and analyze the results, yielding surprising observations about context-sensitivity and
JavaScript. We also briefly evaluate JSAI as compared to
TAJS~\cite{Jensen2009}, the most comparable existing JavaScript
analysis. \ignore{From our evaluation of JSAI, we conclude that
  \textbf{(1)} JSAI's configurability is a useful tool for
  experimenting with JavaScript analysis; and \textbf{(2)} JSAI
  provides useful results for real-world JavaScript applications in a
  reasonable amount of time.}

\ignore{
  \subsection{Implementation} We implemented JSAI in Scala 2.10. In
  addition to the configurable numeric and string domains and the
  tunable control flow, we implemented the following optimizations, to
  improve the interpreter's precision and performance:

  \begin{itemize}
  \renewcommand{\labelitemi }{\ }

  \item \textbf{Type refinement.} The analyzer interprets implicit
  branches that rely on type information (e.g., a conditional branch
  to an exception) and refines the type information that flows into
  the branches by removing facts that are definitely false.

  \item \textbf{Pruning.} Before the interpreter enters a JavaScript
  function, it first removes any locations from the store that are not
  reachable inside the call.

  \item \textbf{Light-weight abstract garbage collection.} Before the
  interpreter exits a function, it first removes from the store any
  strong addresses that correspond to non-escaping local variables. We
  found that this low-cost technique led to reasonable precision and
  performance increases, especially compared to the cost of full
  abstract garbage collection.
  \end{itemize}
}

\subsection{Implementation and Methodology} 

We implement JSAI using Scala version 2.10. 
We provide a model of the DOM, event handling loop, and other APIs used in our benchmarks.
\ignore{The abstract
  interpreter uses the abstract domains described in
  Section~\ref{ssec:domains}. It also employs several optimizations
  not described in this paper, including store pruning, lightweight
  abstract garbage collection, and type refinement. The implementation
  includes a set of context-sensitivities as described in
  Section~\ref{sec:sensitivity}.} The baseline analysis sensitivity we
evaluate is \fs (flow-sensitive, context-insensitive); all of the
other evaluated sensitivities are strictly more precise than \fs. The
other sensitivities are: \stack{$k$}{$h$}, \acyc{$h$},
\objs{$k$}{$h$}, \sig{$k$}{$h$}, and \mixed{$k$}{$h$}, where $k$ is
the context depth for $k$-limiting and $h$ is the heap-sensitivity
(i.e., the prefix of the context used to distinguish abstract
addresses). The parameters $k$ and $h$ vary from 1 to 5 and $h \le k$,
because the heap sensitivity is always a prefix of the context
sensitivity.

We use a comprehensive benchmark suite to evaluate the
sensitivities. Most prior work on JavaScript static analysis has been
evaluated only on the standard SunSpider~\cite{SunSpider} and
V8~\cite{V8bench} benchmarks, with a few micro-benchmarks thrown
in. We evaluate JSAI on these standard benchmarks, but we also include
real-world representatives from a diverse set of JavaScript
application domains. We choose seven representative programs from each
domain, for a total of 28 programs. We partition the programs into
four categories, described below. For each category, we provide the
mean size of the benchmarks in the suite (expressed as number of AST
nodes generated by the Rhino parser~\cite{Rhino}) and the mean
translator blowup (i.e., the factor by which the number of AST nodes
increases when translating from JavaScript to \notjs). The benchmark
names are shown in the graphs presented below; the benchmark suite is
included in the supplementary material. 

The benchmark categories are: \stdbm: seven of the large, complex
benchmarks from SunSpider~\cite{SunSpider} and V8~\cite{V8bench}
(\emph{mean size: 2858 nodes; mean blowup: 8$\times$}); \adnbm: seven
Firefox browser addons, selected from the official Mozilla addon
repository \cite{AMO} (\emph{mean size: 2597 nodes; mean blowup:
  6$\times$}); \emsbm: seven programs from the Emscripten LLVM test
suite, which translates LLVM bitcode to JavaScript~\cite{Emscripten}
(\emph{mean size: 38211 nodes; mean blowup: 7$\times$}); and finally
\opnbm: seven real-world JavaScript programs taken from open source
JavaScript frameworks and their test suites~\cite{Linq, DefensiveJS}
(\emph{mean size: 8784 nodes; mean blowup: 6.4$\times$}).

Our goal is to evaluate the precision and performance of JSAI
instantiated with several forms of context sensitivity. However, the
different sensitivities yield differing sets of function contexts and
abstract addresses, making a fair comparison difficult. Therefore,
rather than statistical measurements (such as address-set size or
closure-set size), we choose a \textit{client-based} precision metric based on a error reporting client. This metric is a proxy for the precision of the analysis.

Our precision metric reports the number
of static program locations (i.e., AST nodes) that might throw
exceptions, based on the analysis' ability to precisely track types.
JavaScript throws a \lstinline|TypeError| exception when a program attempts to
call a non-function or when a program tries to access, update, or
delete a property of \kw{null} or \kw{undef}. JavaScript throws a
\lstinline|RangeError| exception when a program attempts to update the
\lstinline|length| property of an array to contain a value that is not
an unsigned 32-bit integer. Fewer errors indicate a more precise
analysis.

The performance metric we use is execution time of the analysis. To gather data on execution time, we run each
experimental configuration 11 times, discard the first result, then
report the median of the remaining 10 trials. We set a time limit of
30 minutes for each run, reporting a timeout if execution time exceeds that
threshold. We run all experiments on Amazon Web Services~\cite{AWS}
(AWS), using M1 XLarge instances; each experiment is run on an independent AWS instance. These instances have 15GB memory and 8 ECUs, where each ECU is equivalent CPU capacity of
a 1.0-1.2 GHz 2007 Opteron or 2007 Xeon processor.

We run all 56 analyses on each of the 28 benchmarks, for a total
of 1,568 trials (plus the additional 10 executions of each
analysis/benchmark pair for the timing data). For reasons of space, we
present only highlights of these results. In some cases, we present
illustrative examples; the omitted results show similar behavior. In
other cases, we deliberately cherry-pick, to highlight contrasts. We
are explicit about our approach in each case.

\begin{figure*}
\begin{tabular}{p{.30\textwidth}p{.15\textwidth}p{.30\textwidth}}
\multicolumn{3}{r}{
	\includegraphics[scale=1.5]{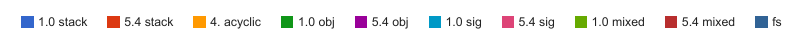}
} \\

\subfloat[\adnbm benchmarks]{
	\includegraphics[scale=0.55]{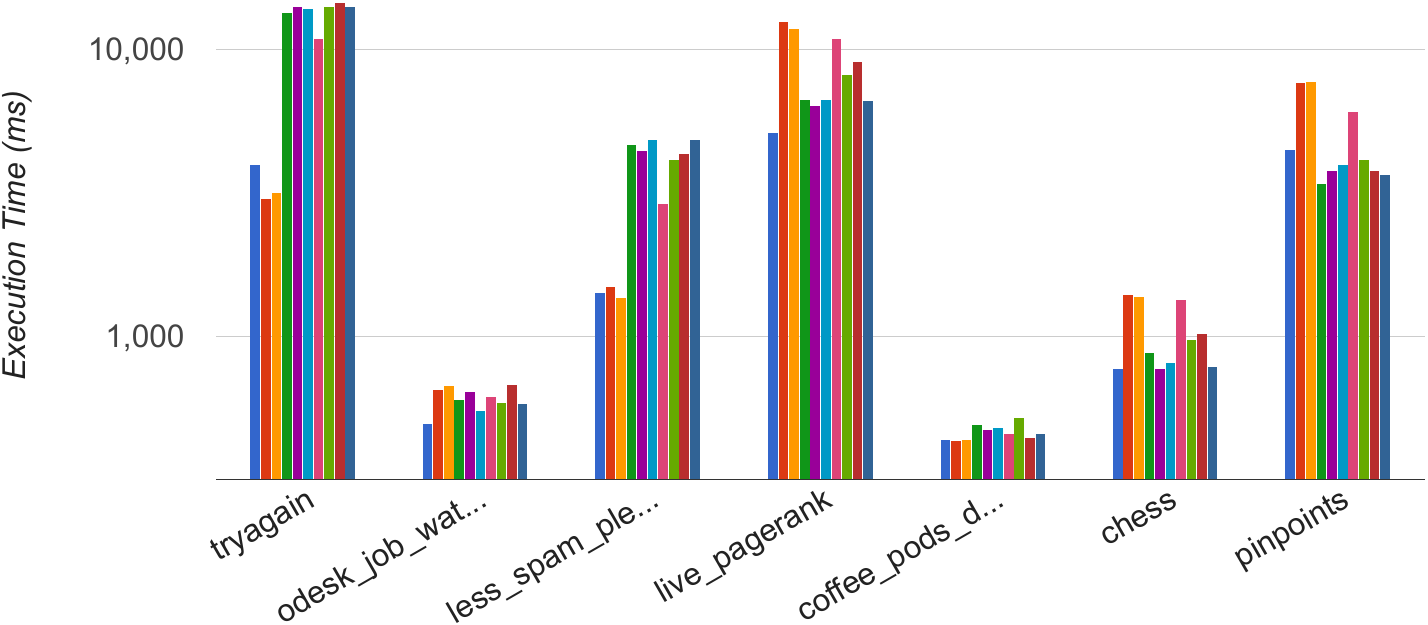}
}& &
\subfloat[\emsbm benchmarks]{
	\includegraphics[scale=0.55]{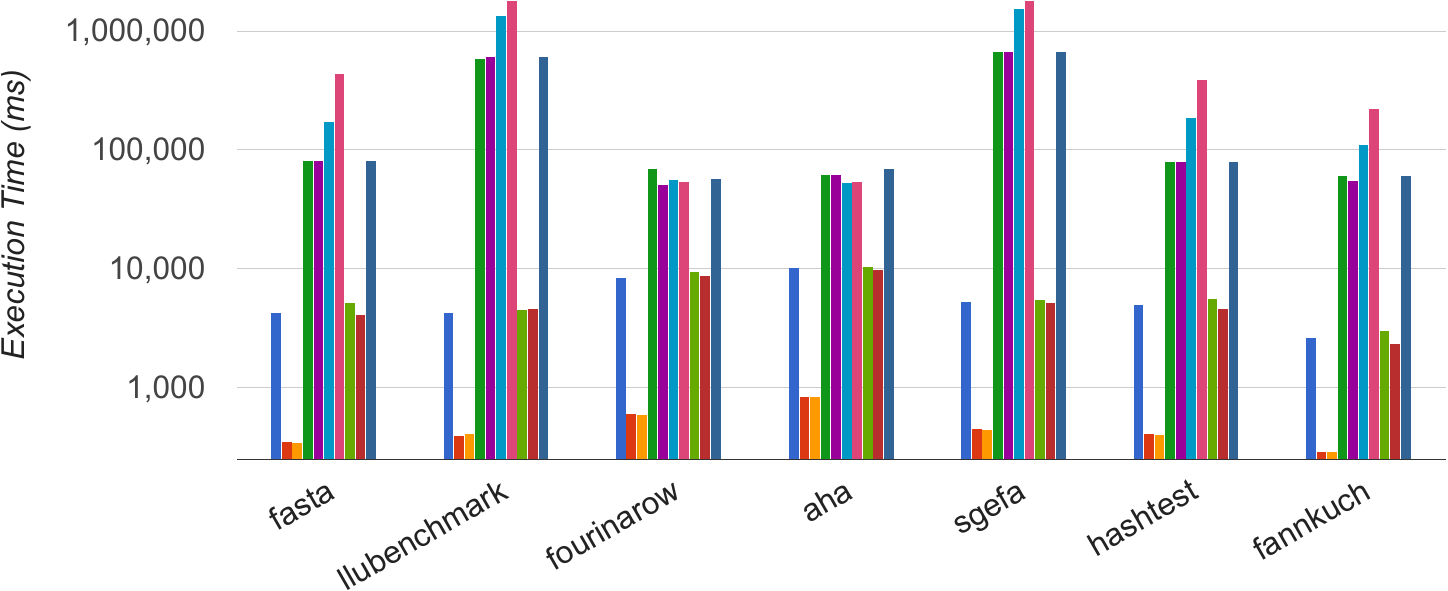}
}\\
\subfloat[\opnbm benchmarks]{
	\includegraphics[scale=0.55]{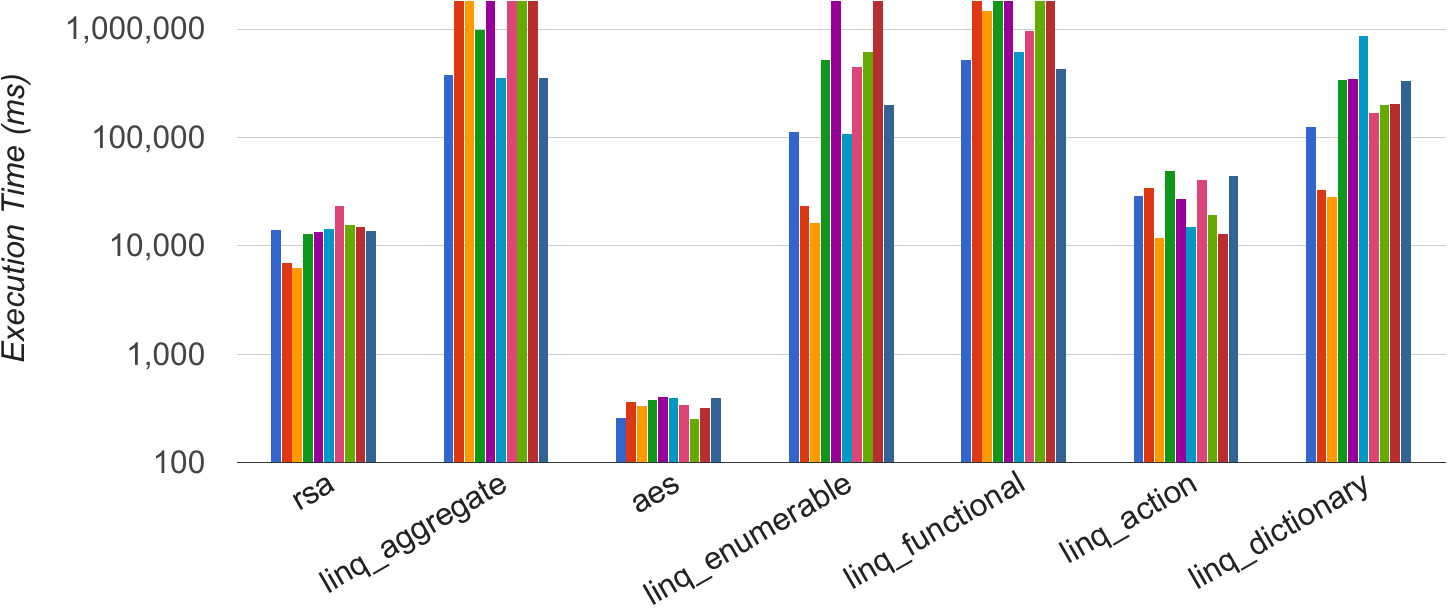}
}& &
\subfloat[\stdbm benchmarks]{
	\includegraphics[scale=0.55]{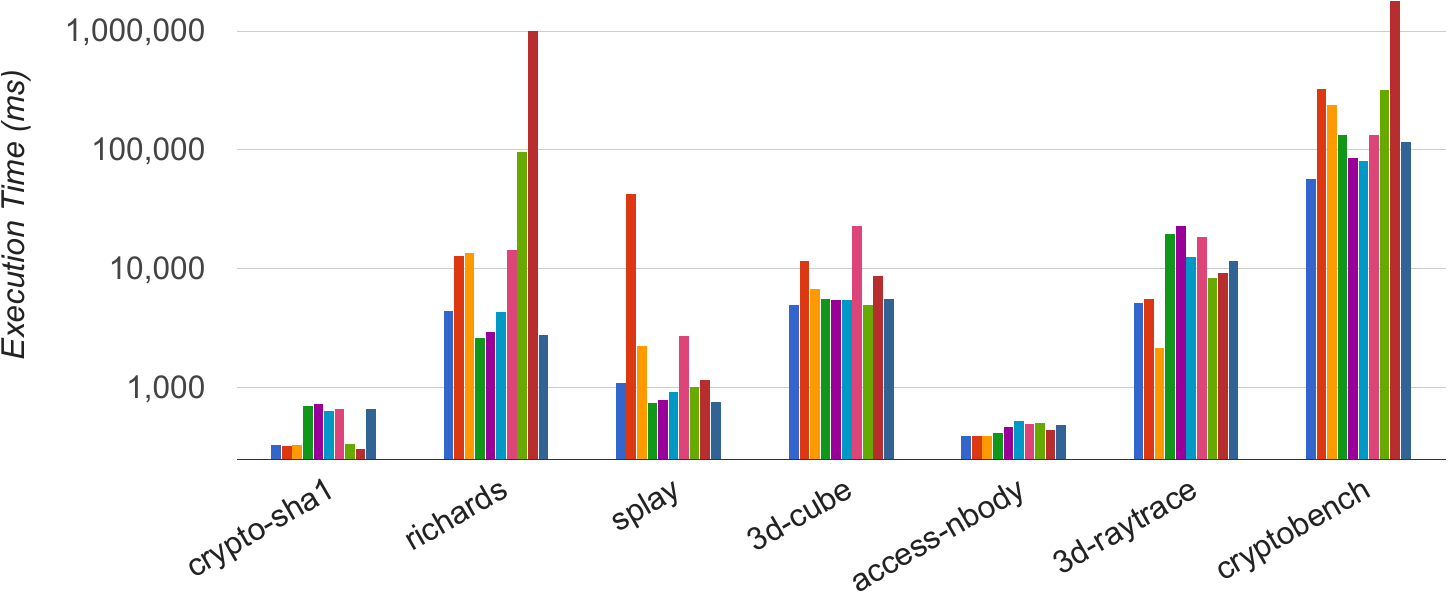} 
}
\end{tabular}
\caption{Performance characteristics of different sensitivities across
the benchmark categories. The x-axis gives the benchmark names. The
y-axis (log scale) gives for each benchmark, the time taken by the analysis (in milliseconds) when run under 10 different sensitivities. Lower
bars mean better performance. Timeout (30 minutes) bars are flush with the top of the graph.} \label{fig:allperf} 
\end{figure*}

\begin{figure}[ht]
\includegraphics[scale=0.5]{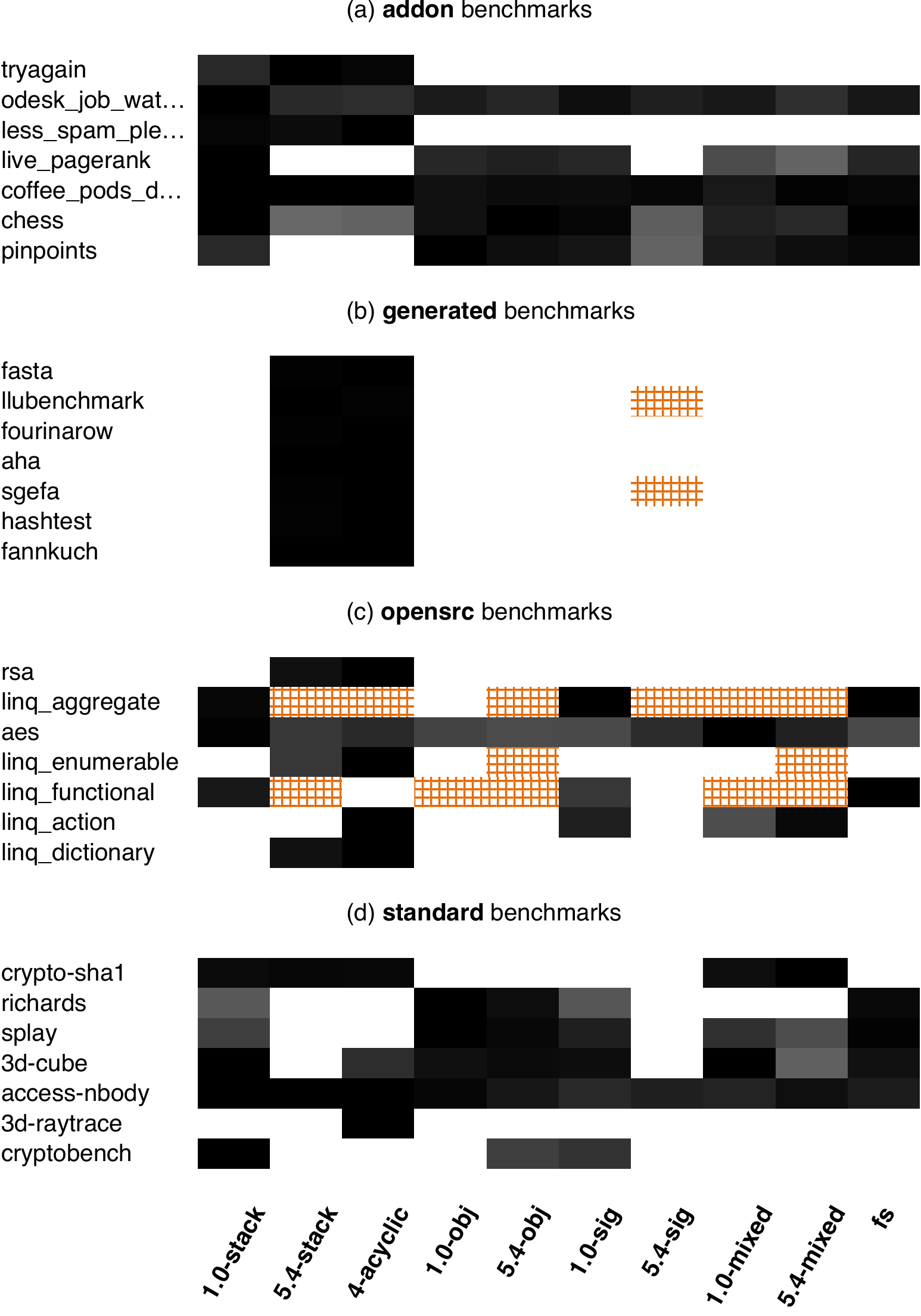}
\caption{
A heat map to showcase the performance characteristics of different sensitivities across the benchmark categories.
The above figure is a two-dimensional map of blocks; rows correspond to benchmarks, and columns correspond to analysis run with a particular sensitivity. 
The color in a block indicates a sensitivities' relative performance on the corresponding benchmark, as compared to fastest sensitivity on that benchmark. 
Darker colors represent better performance.
Completely blackened blocks indicate that the corresponding sensitivity has the fastest analysis time on that benchmark, while completely whitened blocks indicate that the corresponding sensitivity does not time out, but has a relative slowdown of at least $2 \times$. 
The remaining colors are of evenly decreasing contrast from black to white, representing a slowdown between $1 \times$ to $2 \times$.
The red grid pattern on a block indicates a timeout. 
}

\label{fig:performance-heatmap}
\end{figure}

\begin{figure}[ht]
\includegraphics[scale=0.5]{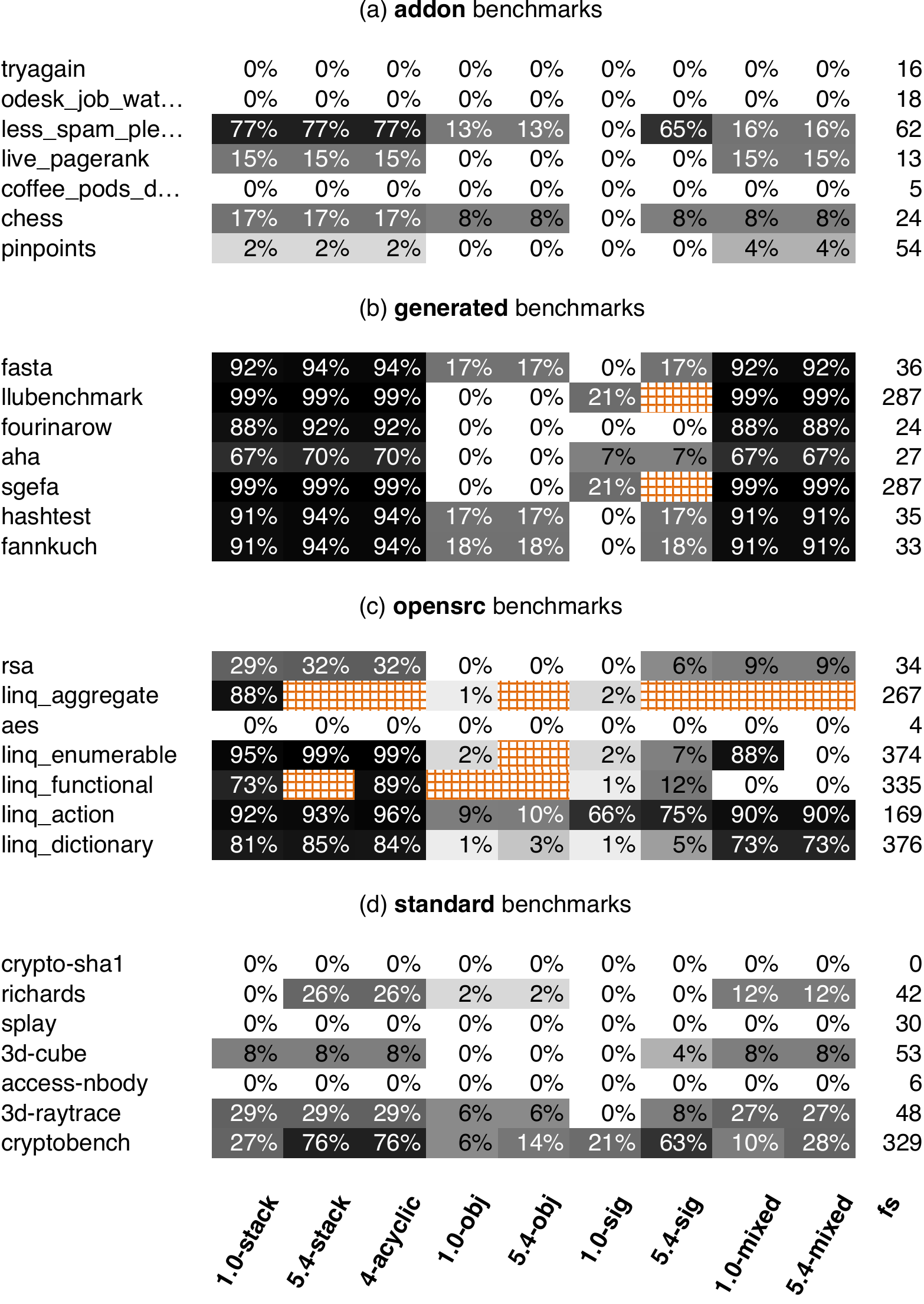}
\caption{A heat map to showcase the precision characteristics (based on number of reported runtime errors) of different sensitivities across the benchmark categories.
The above figure is a two-dimensional map of blocks; rows correspond to benchmarks, and columns corresponds to analysis run with a particular sensitivity.
The rightmost column corresponds to the context insensitive analysis \fs, and the blocks in this column give the number of errors reported by the analysis under \fs (which is an upper bound on the number of errors reported across any sensitivity).
The color (which ranges evenly from black to white) in the remaining blocks indicate the percentage reduction in number of errors reported by the analysis under the corresponding sensitivity, compared to \fs on the same benchmark.
Darker colors represent more reduction in errors reported, and hence better precision.
In addition to the colors, the percentage reduction in errors is also given inside the blocks (higher percentage reduction indicates better precision).
The red grid pattern on a block indicates a timeout. 
}
\label{fig:precision-heatmap}
\end{figure}

\subsection{Observations}

For each main sensitivity strategy, we present the data for two
trials: the least precise sensitivity in that strategy, and the most
precise sensitivity in that strategy. This set of
analyses
is: \fs, \stack{1}{0}, \stack{5}{4}, \acyc{4}, \objs{1}{0}, \objs{5}{4},
\sig{1}{0}, \sig{5}{4}, \mixed{1}{0}, \mixed{5}{4}. 

Figures~\ref{fig:allperf} and~\ref{fig:performance-heatmap} contain performance results, and Figure ~\ref{fig:precision-heatmap} contains the precision results. 
The results are partitioned by benchmark category to show the effect of each analysis sensitivity on benchmarks in that category. 
The performance graphs in Figure~\ref{fig:allperf} plot the median
execution time in milliseconds, on a log scale, giving a sense of actual time taken by the various sensitivity strategies. Lower bars are better; timeouts extend above the top of the graph. 

We provide an alternate visualization of the performance data through Figure~\ref{fig:performance-heatmap} to easily depict how the sensitivities perform relative to each other.
Figure~\ref{fig:performance-heatmap} is heat map that lays out blocks in two dimensions---rows represent benchmarks and columns represent analyses with different sensitivities.
Each block represents relative performance as a color: darker blocks correspond to faster execution time of a sensitivity compared to other sensitivities on the same benchmark.
A completely blackened block corresponds to the fastest sensitivity on that benchmark, a whitened block corresponds to a sensitivity that has $\ge 2 \times$ slowdown relative to the fastest sensitivity, and the remaining colors evenly correspond to slowdowns in between.
Blocks with the red grid pattern indicate a timeout.
A visual cue is that columns with darker blocks correspond to better-performing sensitivities, and a row with blocks that have very similar colors indicates a benchmark on which performance is unaffected by varying sensitivities. 

Figure~\ref{fig:precision-heatmap} provides a similar heat map (with similar visual cues) for visualizing relative precisions of various sensitivity strategies on our benchmarks.
The final column in this heat map provides the number of errors reported by the \fs strategy on a particular benchmark, while the rest of the columns provide the percentage reduction (relative to \fs) in the number of reported errors due to a corresponding sensitivity strategy.
The various blocks (except the ones in the final column) are color coded in addition to providing percentage reduction numbers: darker is better precision (that is, more reduction in number of reported errors).
Timeouts are indicated using a red grid pattern.

\paragraph{\textit{Breaking the Glass Ceiling.}}

\begin{figure}
\centering
\includegraphics[scale=0.55]{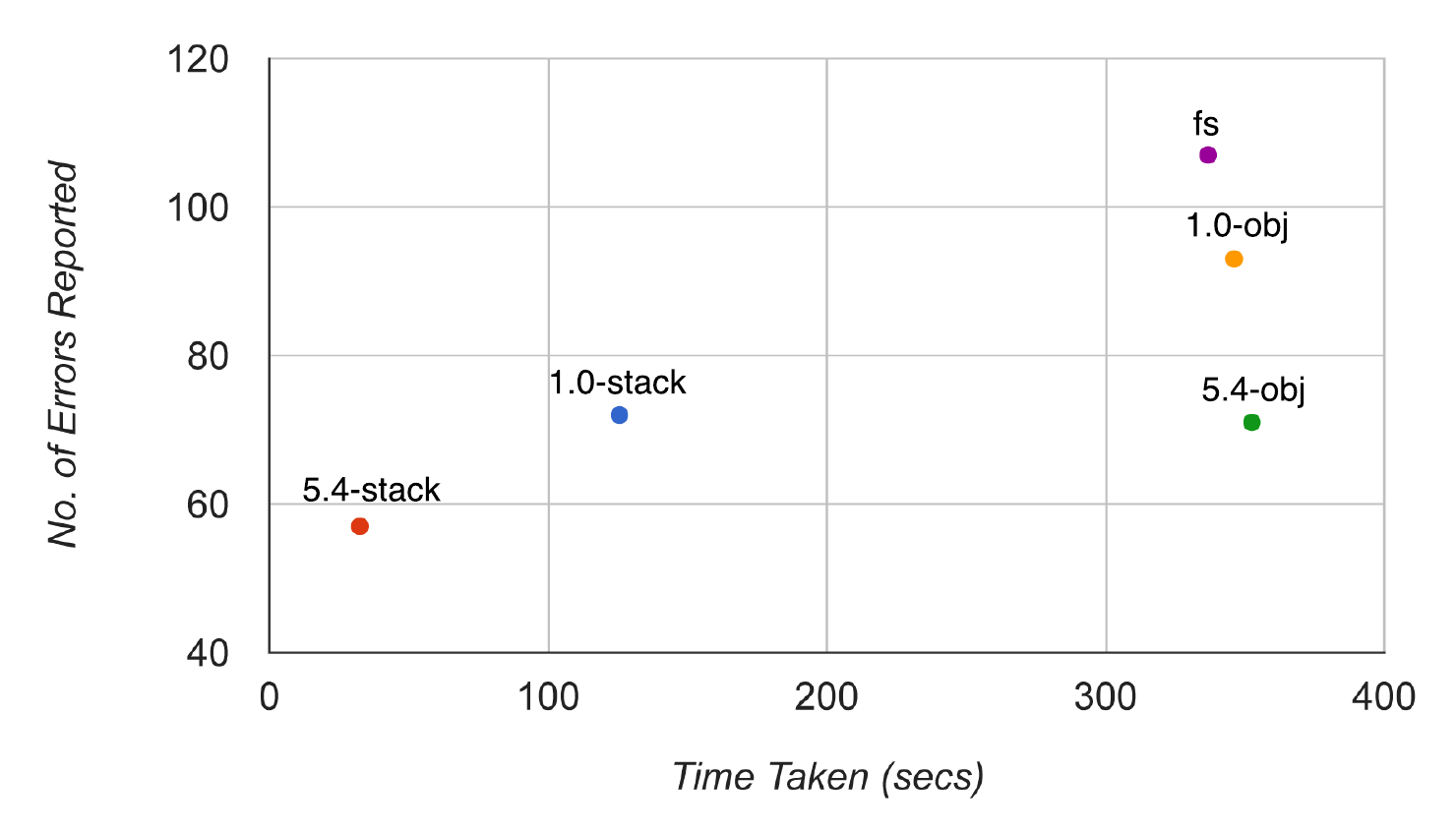}
\caption{Precision \emph{vs.} performance of various sensitivities, on
the \opnbm \ttt{linq\_dictionary} benchmark.
Interestingly, \stack{5}{4} (the most sensitive Stack-CFA analysis) is not only tractable, it exhibits the best performance and the best precision.}
\label{fig:scatter}
\end{figure}

One startling observation is that highly sensitive variants (i.e.,
sensitivity strategies with high $k$ and $h$ parameters) can be far
better than their less-sensitive counterparts, providing improved
precision at a much cheaper cost (see Figure~\ref{fig:scatter}). For example, on
\ttt{linq\_dictionary}, \stack{5}{4} is the most precise \emph{and}
most efficient analysis. By contrast, the \stack{3}{2} analysis yields
the same result at a three-fold increase in cost, while
the \stack{1}{0} analysis is even more expensive and less precise. 
We see similar behavior for the \ttt{sgefa} benchmark, where \stack{5}{4} is an order of magnitude faster than \stack{1}{0} and delivers the same results. This behavior violates the common wisdom that values of
$k$ and $h$ above 1 or 2 are intractably expensive.

This behavior is certainly not universal,\footnote{For example,
\ttt{linq\_aggregate} times out on all analyses with $k>1$.} but it is
intriguing. Analysis designers often try to scale up their
context-sensitivity (in terms of $k$ and $h$) linearly, and they stop
when it becomes intractable. However, our experiments suggest that
pushing past this local barrier may yield much better results.

\paragraph{\textit{Callstring vs Object Sensitivity.}}
In general, we find that callstring-based sensitivity
(i.e., \stack{$k$}{$h$} and \acyc{$h$}) is more precise than object
sensitivity (i.e., \objs{$k$}{$h$}). This result is unintuitive, since
JavaScript heavily relies on objects and object sensitivity was
specifically designed for object-oriented languages such as Java.
Throughout the benchmarks, the most precise and efficient analyses are
the ones that employ stack-based $k$-CFA. Part of the reason for this
trend is that 25\% of the benchmarks are machine-generated JavaScript
versions of procedural code, whose structure yields more benefits to
callstring-based context-sensitivity. Even among the handwritten open-source
benchmarks, however, this trend holds. For
example, several forms of callstring sensitivity are more efficient
and provide more precise results for the open-source benchmarks than object-sensitivity,
which often times out. \ignore{There are a few benchmarks for which the most
precise analysis is object-sensitive, but most of these benchmarks
(e.g., \ttt{coffee\_pods\_deals}) exhibit little variance in their
precision. Based on these observations, it seems wise to use
object-sensitivity only after confirming that stack-based $k$-CFA
cannot provide enough precision or performance for a particular
program or suite of programs.}

\ignore{
  % BAW: removed for space because I think it emphasizes a 
  % point we already make (rather than illustrates a new point)
  Consider the Table~\ref{tbl:sgefa}, which provides the performance
  and error metric precision information for various analyses run
  under various sensitivities, for the benchmark \ttt{sgefa} in EMS
  category. It can be seen that the stack and acyclic sensitivities do
  really well for this benchmark, and the others do very poorly (for
  both performance and precision). This is a recurring theme
  throughout the EMS benchmark category.

  \input{figures/sgefa.tex}
}

\paragraph{\textit{Benefits of Context Sensitivity.}}
When it comes to pure precision, we find that more context sensitivity
sometimes increases precision and sometimes has no effect.
The open-source benchmarks demonstrate quite a bit of variance for the precision metric. 
A context-sensitive analysis almost always finds fewer errors (i.e.,
fewer false positives) than a context-insensitive analysis, and
increasing the sensitivity in a particular family leads to precision
gains. For example, \stack{5}{4} gives the most precise error report
for \ttt{linq\_enumerable}, and it is an order of magnitude more precise
than a context-insensitive analysis. 
\ignore{Data dependence does not seem to
benefit much from increased sensitivity. This behavior is most
apparent in the \adnbm benchmarks, where all the sensitivities find
roughly the same number of dependencies. It may be that this metric is
not context-sensitive: that there are not a lot of interprocedural
dependences conditioned on the interprocedural control flow.
Based on
these observations, it might be wise to focus on increasing precision
via other means, e.g., via more strong updates or more precise
abstract domains.}
The addon domain has very little variance for the precision metric, which is perhaps due to shorter call sequence lengths in this domain. 
In such domains, it might be wise to focus on performance, rather than increasing precision.

\ignore{ \note{maybe this paragraph belongs in Future Work?} One
  technique that may improve performance is sparseness. A sparse
  analysis directly connects the producers of analysis facts to their
  consumers, bypassing intermediate nodes or states that act as mere
  carriers. A sparse analysis has the potential to give huge
  performance gains, but sparse analysis requires precise control-flow
  information, itself a product of analysis. Doing staged analysis,
  which uses information from a cheaper analysis is also difficult: as
  we have seen, less precise does not mean cheap in terms of
  performance. While lazy propagation~\cite{Jensen2010} is a step in
  the direction of bringing sparse analysis to JavaScript, it still is
  an open research question.

  We also observe that our novel mixed and signature-based
  sensitivities behaves similarly to object sensitivities. We had
  thought that the ideas behind these seemed intuitive and expected
  them to show marked improvements over standard object or
  callstring-based sensitivities, but they did not.  Nevertheless, it
  was easy to implement our ideas in a few lines of code (without
  modifying the rest of the analysis). Thus we see the result as a
  validation of JSAI as an experimental platform that lets us quickly
  try out new ideas on real JavaScript programs.  }

  \paragraph{\textit{Summary.}}
  Perhaps the most sweeping claim we can make from the data is that
  there is no clear winner across all benchmarks, in terms of JavaScript
  context-sensitivity. This state of affairs is not a surprise: the
  application domains for JavaScript are so rich and varied that
  finding a silver bullet for precision and performance is
  unlikely. However, it is likely that---within an application domain,
  e.g., automatically generated JavaScript code---one form of
  context-sensitivity could emerge a clear winner. The benefit of JSAI
  is that it is easy to experiment with the context-sensitivity of an
  analysis. The analysis designer need only implement their base
  analysis, then instantiate and evaluate multiple instances of the
  analysis to tune context-sensitivity.

\subsection{Discussion: JSAI \emph{vs.} TAJS} 
\label{sec:tajs}

Jensen \etal's Type Analysis for JavaScript~\cite{Jensen2009, Jensen2010} (TAJS) stands
as the only published static analysis for JavaScript whose intention
is to soundly analyze the entire JavaScript language. JSAI has several
features that TAJS does not, including configurable sensitivity, a
formalized abstract semantics, and novel abstract domains, but TAJS is
a valuable contribution that has been put to good use. An interesting
question is how JSAI compares to TAJS in terms of precision and
performance.

The TAJS implementation (in Java) has matured over a period of five years, it
has been heavily optimized, and it is publicly available. Ideally, we
could directly compare TAJS to JSAI with respect to precision and
performance, but they are dissimilar enough that they are effectively
noncomparable. For one, TAJS has known soundness bugs\footnote{We
  uncovered several soundness bugs when we were formalizing our semantics, and
  the TAJS authors confirmed them as errors.} that can artificially
decrease its set of reported type errors. Also, TAJS does not
implement some of the APIs required by our benchmark suite, and so it can
only run on a subset of the benchmarks. On the flip side, TAJS
is more mature than JSAI, it has a more precise implementation of the
core JavaScript APIs, and it contains a number of precision and
performance optimizations (e.g., the recency heap
abstraction~\cite{Balakrishnan2006} and lazy propagation
\cite{Jensen2010}) that JSAI does not currently implement.

Nevertheless, we can perform a qualitative ``ballpark" comparison, to
demonstrate that JSAI is roughly comparable in terms of precision and
performance. For the subset of our benchmarks on which both JSAI and
TAJS execute, we catalogue the number of errors that each tool
reports and record the time it took for each tool to do so. We find
that JSAI analysis time is 0.3$\times$ to 1.8$\times$ that of TAJS. In
terms of precision, JSAI reports from nine fewer type errors to 104
more type errors, compared to TAJS. Many of the extra type errors that
JSAI reports are \lstinline|RangeError|s, which TAJS does not report
due to one of the unsoundness bugs we uncovered. Excluding
\lstinline|RangeError|s, JSAI reports at most 20 more errors than TAJS
in the worst case.

\section{Conclusion and Future Work}
\label{sec:conclusion}

We have described the design of JSAI, a configurable, sound, and \practical abstract interpreter for JavaScript. 
JSAI's design is novel in a number of respects which make it stand out from all previous JavaScript analyzers. 
We have provided a comprehensive evaluation that demonstrates JSAI's usefulness. The JSAI implementation and formalisms are freely available as a supplement, and we believe that JSAI will provide a useful platform for researchers investigating JavaScript analysis.

Our future work includes (1) taking advantage of JSAI's tunable precision to further investigate what control-flow sensitivities are most useful for JavaScript; (2) writing a number of clients on top of JSAI, including program refactoring, program compression; and (3) extending JSAI to handle language features from the latest ECMA 5
standard.
\balance

%\setstretch{1}
\bibliographystyle{abbrvnat}
\bibliography{paper}

\begin{thebibliography}{48}
\providecommand{\natexlab}[1]{#1}
\providecommand{\url}[1]{\texttt{#1}}
\expandafter\ifx\csname urlstyle\endcsname\relax
  \providecommand{\doi}[1]{doi: #1}\else
  \providecommand{\doi}{doi: \begingroup \urlstyle{rm}\Url}\fi

\bibitem[AMO()]{AMO}
\url{https://addons.mozilla.org/en-US/firefox/}.

\bibitem[AWS()]{AWS}
\url{http://aws.amazon.com/}.

\bibitem[Def()]{DefensiveJS}
\url{http://www.defensivejs.com/}.

\bibitem[Doc()]{DoctorJS}
\url{http://doctorjs.org/}.

\bibitem[Ems()]{Emscripten}
\url{http://www.emscripten.org/}.

\bibitem[Esp()]{Esposito2012}
\url{http://www.drdobbs.com/windows/microsofts-javascript-move/240012790}.

\bibitem[Fir()]{FirefoxOS}
\url{http://www.mozilla.org/en-US/firefox/os/}.

\bibitem[Lin()]{Linq}
\url{http://linqjs.codeplex.com/}.

\bibitem[Rhi()]{Rhino}
\url{https://developer.mozilla.org/en-US/docs/Rhino}.

\bibitem[Sta()]{StarToJS}
\url{https://github.com/jashkenas/coffee-script/wiki/List-of-languages-that-compile-to-JS}.

\bibitem[Sun()]{SunSpider}
\url{http://www.webkit.org/perf/sunspider/sunspider.html}.

\bibitem[V8b()]{V8bench}
\url{http://v8.googlecode.com/svn/data/benchmarks/v7/run.html}.

\bibitem[nod()]{nodejs}
\url{http://nodejs.org/}.

\bibitem[spi()]{spidermonkey}
\url{https://developer.mozilla.org/en-US/docs/SpiderMonkey}.

\bibitem[typ()]{typedarray}
\url{http://www.khronos.org/registry/typedarray/specs/latest/}.

\bibitem[Anderson et~al.(2005)Anderson, Giannini, and
  Drossopoulou]{Anderson2005}
C.~Anderson, P.~Giannini, and S.~Drossopoulou.
\newblock Towards type inference for javascript.
\newblock In \emph{European conference on Object-oriented programming}, 2005.

\bibitem[Balakrishnan and Reps(2006)]{Balakrishnan2006}
G.~Balakrishnan and T.~Reps.
\newblock Recency-abstraction for heap-allocated storage.
\newblock In \emph{International conference on Static Analysis}, 2006.

\bibitem[Bandhakavi et~al.(2011)Bandhakavi, Tiku, Pittman, King, Madhusudan,
  and Winslett]{Bandhakavi2011}
S.~Bandhakavi, N.~Tiku, W.~Pittman, S.~T. King, P.~Madhusudan, and M.~Winslett.
\newblock Vetting browser extensions for security vulnerabilities with vex.
\newblock \emph{Commun. ACM}, 54\penalty0 (9), Sept. 2011.

\bibitem[Chugh et~al.(2009)Chugh, Meister, Jhala, and Lerner]{Chugh2009}
R.~Chugh, J.~A. Meister, R.~Jhala, and S.~Lerner.
\newblock Staged information flow for javascript.
\newblock In \emph{ACM SIGPLAN Conference on Programming Languages Design and
  Implementation}, 2009.

\bibitem[Chugh et~al.(2012)Chugh, Herman, and Jhala]{Chugh2012}
R.~Chugh, D.~Herman, and R.~Jhala.
\newblock Dependent types for javascript.
\newblock In \emph{International Conference on Object Oriented Programming
  Systems Languages and Applications}, 2012.

\bibitem[Cousot and Cousot(1977)]{Cousot1977}
P.~Cousot and R.~Cousot.
\newblock Abstract interpretation: a unified lattice model for static analysis
  of programs by construction or approximation of fixpoints.
\newblock In \emph{ACM Symposium on Principles of programming languages}. ACM
  Press, New York, NY, 1977.

\bibitem[Cousot and Cousot(1979)]{Cousot1979}
P.~Cousot and R.~Cousot.
\newblock Systematic design of program analysis frameworks.
\newblock In \emph{ACM Symposium on Principles of Programming Languages}, 1979.

\bibitem[{ECMA}(1999)]{ECMA-262}
{ECMA}.
\newblock \emph{{ECMA-262}: {ECMAScript} Language Specification}.
\newblock Third edition, Dec. 1999.
\newblock URL \url{http://www.ecma.ch/ecma1/STAND/ECMA-262.HTM}.

\bibitem[Gardner et~al.(2012)Gardner, Maffeis, and Smith]{Gardner2012}
P.~A. Gardner, S.~Maffeis, and G.~D. Smith.
\newblock Towards a program logic for javascript.
\newblock In \emph{ACM Symposium on Principles of programming languages}, 2012.

\bibitem[Guarnieri and Livshits(2009)]{Guarnieri2009}
S.~Guarnieri and B.~Livshits.
\newblock Gatekeeper: mostly static enforcement of security and reliability
  policies for javascript code.
\newblock In \emph{Conference on USENIX security symposium}, 2009.

\bibitem[Guha et~al.(2009)Guha, Krishnamurthi, and Jim]{Guha2009}
A.~Guha, S.~Krishnamurthi, and T.~Jim.
\newblock Using static analysis for {Ajax} intrusion detection.
\newblock In \emph{World Wide Web Conference}, 2009.

\bibitem[Guha et~al.(2010)Guha, Saftoiu, and Krishnamurthi]{Guha2010}
A.~Guha, C.~Saftoiu, and S.~Krishnamurthi.
\newblock The essence of javascript.
\newblock In \emph{European conference on Object-oriented programming}, 2010.

\bibitem[Guha et~al.(2011)Guha, Saftoiu, and Krishnamurthi]{Guha2011}
A.~Guha, C.~Saftoiu, and S.~Krishnamurthi.
\newblock Typing local control and state using flow analysis.
\newblock In \emph{European conference on Programming languages and systems},
  2011.

\bibitem[Hardekopf et~al.(2014)Hardekopf, Wiedermann, Churchill, and
  Kashyap]{Hardekopf2014}
B.~Hardekopf, B.~Wiedermann, B.~Churchill, and V.~Kashyap.
\newblock Widening for control-flow.
\newblock In \emph{International Conference on Verification, Model Checking,
  and Abstract Interpretation}, 2014.

\bibitem[Heidegger and Thiemann(2010)]{Heidegger2010}
P.~Heidegger and P.~Thiemann.
\newblock Recency types for analyzing scripting languages.
\newblock \emph{European conference on Object-oriented programming}, 2010.

\bibitem[Ishizaki et~al.(2000)Ishizaki, Kawahito, Yasue, Komatsu, and
  Nakatani]{Ishizaki2000}
K.~Ishizaki, M.~Kawahito, T.~Yasue, H.~Komatsu, and T.~Nakatani.
\newblock A study of devirtualization techniques for a java just-in-time
  compiler.
\newblock In \emph{ACM International Conference on Object Oriented Programming
  Systems Languages and Applications}, 2000.

\bibitem[Jang and Choe(2009)]{Jang2009}
D.~Jang and K.-M. Choe.
\newblock Points-to analysis for javascript.
\newblock In \emph{Symposium on Applied Computing}, 2009.

\bibitem[Jensen et~al.(2009)Jensen, M{\o}ller, and Thiemann]{Jensen2009}
S.~H. Jensen, A.~M{\o}ller, and P.~Thiemann.
\newblock {T}ype {A}nalysis for {J}avascript.
\newblock In \emph{International Symposium on Static Analysis}, 2009.

\bibitem[Jensen et~al.(2010)Jensen, M\o{}ller, and Thiemann]{Jensen2010}
S.~H. Jensen, A.~M\o{}ller, and P.~Thiemann.
\newblock {I}nterprocedural {A}nalysis with {L}azy {P}ropagation.
\newblock In \emph{International Symposium on Static Analysis}, 2010.

\bibitem[Jensen et~al.(2012)Jensen, Jonsson, and M\o{}ller]{Jensen2012}
S.~H. Jensen, P.~A. Jonsson, and A.~M\o{}ller.
\newblock {R}emedying the {E}val that {M}en {D}o.
\newblock In \emph{International Symposium on Software Testing and Analysis},
  2012.

\bibitem[Kashyap and Hardekopf(2014)]{Kashyap2014}
V.~Kashyap and B.~Hardekopf.
\newblock Security signature inference for javascript-based browser addons.
\newblock In \emph{Symposium on Code Generation and Optimization}, 2014.

\bibitem[Kashyap et~al.(2013)Kashyap, Sarracino, Wagner, Wiedermann, and
  Hardekopf]{Kashyap2013}
V.~Kashyap, J.~Sarracino, J.~Wagner, B.~Wiedermann, and B.~Hardekopf.
\newblock Type refinement for static analysis of javascript.
\newblock In \emph{Symposium on Dynamic Languages}, 2013.

\bibitem[Lee et~al.(2012)Lee, Won, Jin, Cho, and Ryu]{Lee2012}
H.~Lee, S.~Won, J.~Jin, J.~Cho, and S.~Ryu.
\newblock Safe: Formal specification and implementation of a scalable analysis
  framework for ecmascript.
\newblock In \emph{International Workshop on Foundations of Object-Oriented
  Languages}, 2012.

\bibitem[Logozzo and Venter(2010)]{Logozzo2010}
F.~Logozzo and H.~Venter.
\newblock Rata: {R}apid {A}tomic {T}ype {A}nalysis by {A}bstract
  {I}nterpretation -- {A}pplication to {J}avascript {O}ptimization.
\newblock In \emph{Joint European Conference on Theory and Practice of
  Software, International Conference on Compiler Construction}, 2010.

\bibitem[Madsen et~al.(2013)Madsen, Livshits, and Fanning]{Madsen2013}
M.~Madsen, B.~Livshits, and M.~Fanning.
\newblock Practical static analysis of {JavaScript} applications in the
  presence of frameworks and libraries.
\newblock In \emph{ACM Symposium on the Foundations of Software Engineering},
  Aug. 2013.

\bibitem[Maffeis et~al.(2008)Maffeis, Mitchell, and Taly]{Maffeis2008}
S.~Maffeis, J.~C. Mitchell, and A.~Taly.
\newblock An operational semantics for javascript.
\newblock In \emph{Asian Symposium on Programming Languages and Systems}, 2008.

\bibitem[Meawad et~al.(2012)Meawad, Richards, Morandat, and Vitek]{Meawad2012}
F.~Meawad, G.~Richards, F.~Morandat, and J.~Vitek.
\newblock Eval begone!: semi-automated removal of eval from javascript
  programs.
\newblock In \emph{ACM International Conference on Object Oriented Programming
  Systems Languages and Applications}, 2012.

\bibitem[Sch\"{a}fer et~al.(2013)Sch\"{a}fer, Sridharan, Dolby, and
  Tip]{Schafer2013}
M.~Sch\"{a}fer, M.~Sridharan, J.~Dolby, and F.~Tip.
\newblock Dynamic determinacy analysis.
\newblock In \emph{ACM SIGPLAN Conference on Programming Languages Design and
  Implementation}. ACM, 2013.

\bibitem[Sergey et~al.(2013)Sergey, Devriese, Might, Midtgaard, Darais, Clarke,
  and Piessens]{Sergey2013}
I.~Sergey, D.~Devriese, M.~Might, J.~Midtgaard, D.~Darais, D.~Clarke, and
  F.~Piessens.
\newblock Monadic abstract interpreters.
\newblock In \emph{ACM SIGPLAN Conference on Programming Languages Design and
  Implementation}. ACM, 2013.

\bibitem[Smaragdakis et~al.(2011)Smaragdakis, Bravenboer, and
  Lhot\'{a}k]{Smaragdakis2011}
Y.~Smaragdakis, M.~Bravenboer, and O.~Lhot\'{a}k.
\newblock Pick your contexts well: understanding object-sensitivity.
\newblock In \emph{ACM Symposium on Principles of programming languages}, 2011.

\bibitem[Taly et~al.(2011)Taly, Erlingsson, Mitchell, Miller, and
  Nagra]{Taly2011}
A.~Taly, U.~Erlingsson, J.~C. Mitchell, M.~S. Miller, and J.~Nagra.
\newblock Automated analysis of security-critical javascript apis.
\newblock In \emph{IEEE Symposium on Security and Privacy}, 2011.

\bibitem[Thiemann(2005)]{Thiemann2005}
P.~Thiemann.
\newblock {T}owards a {T}ype {S}ystem for {A}nalyzing {J}avascript {P}rograms.
\newblock In \emph{European Conference on Programming Languages and Systems},
  2005.

\bibitem[Van~Horn and Might(2010)]{VanHorn2010}
D.~Van~Horn and M.~Might.
\newblock Abstracting abstract machines.
\newblock In \emph{International Conference on Functional Programming}, 2010.

\end{thebibliography}

\end{document}